\documentclass{aa}
\usepackage{graphics}

\makeatletter

\begin{document}

\thesaurus{10(02.14.1;08.01.1;10.01.1;10.05.1;10.07.1)}
\title{Galactic chemical abundance evolution in the solar neighborhood up to the Iron
peak}

\author{Andreu Alib\'es\inst{1}
  \and Javier Labay\inst{1}
  \and Ramon Canal\inst{1,2}}

\offprints{A.Alib\'es}

\institute{Departament d'Astronomia i Meteorologia, Universitat de Barcelona,
           {Mart\'\i} i Franqu\`es 1, 08028 Barcelona (Spain)
	   \and Institut d'Estudis Espacials de Catalunya, Edifici Nexus, Gran
	   Capit\`a 2-4, 08034 Barcelona (Spain)\\
	   email: aalibes@am.ub.es}

\date{Received / Accepted}

\authorrunning{A. Alib\'es, J. Labay and R. Canal}
\titlerunning{Solar elemental abundances evolution}

\maketitle
\begin{abstract}
We have developed a detailed standard chemical evolution model to study the
evolution of all the chemical elements up to the iron peak in the solar vicinity.
We consider that the Galaxy was formed through two episodes of exponentially
decreasing infall, out of extragalactic gas. In a first infall episode, with
a duration of \( \sim  \) 1 Gyr, the halo and the thick disk were assembled
out of primordial gas, while the thin disk formed in a second episode of infall
of slightly enriched extragalactic gas, with much longer timescale. The model
nicely reproduces the main observational constraints of the solar neighborhood,
and the calculated elemental abundances at the time of the solar birth are in
excellent agreement with the solar abundances. By the inclusion of metallicity
dependent yields for the whole range of stellar masses we follow the evolution
of 76 isotopes of all the chemical elements between hydrogen and zinc. Those
results are confronted with a large and recent body of observational data, and
we discuss in detail the implications for stellar nucleosynthesis.
\end{abstract}

\section{Introduction}

The chemical evolution of our Galaxy has been extensively studied during the
last years. Along the last decade a great deal of new data have become available,
revealing in principle the chemical history of the Milky Way and building up
a set of observational constraints that have to be fulfilled by any successful
theoretical model. These observational results concern in particular the age-metallicity
relation (Meussinger et al. \cite{meus91}; Edvardsson et al. \cite{edva93};
Rocha-Pinto et al. \cite{roch00}), star metallicity distribution (Wyse \& Gilmore
\cite{wyse95}; Rocha-Pinto \& Maciel \cite{roch96}; J\o rgensen \cite{jorg00}),
abundance ratios of an increasing number of chemical elements, both in halo
and local disk stars (between many others, Sneden et al. \cite{sned94}; McWilliam
et al. \cite{mcwi95}; Ryan et al. \cite{ryan96}; Chen et al. \cite{chen00}),
and radial abundance gradients (see Henry \& Worthey \cite{henr99} for a review).

These observational efforts have been accompanied by the publication of several
theoretical models which try to interpret the data. Most of that work are standard
open chemical evolution models (Carigi \cite{cari94}; Giovanoli \& Tosi \cite{giov95};
Prantzos \& Aubert \cite{pran95}; Timmes et al. \cite{timm95} (hereafter TWW1995);
Chiappini et al. \cite{chia97}, \cite{chia99}; Thomas et al. \cite{thom98};
Chang et al. \cite{chan99}; Goswami \& Prantzos \cite{gosw00} (hereafter GP2000),
etc.) in which the Galaxy is assembled by infall of extragalactic gas and that
use different prescriptions for the main ingredients, i.e. infall rate and composition
of the accreted material, initial mass function (IMF), star formation rate (SFR)
and stellar yields. Chemodynamical models have also been published (Steinmetz
\& Müller \cite{stei94}; Samland et al. \cite{saml97}; Samland \cite{saml98},
Berczik \cite{berc99}), but due to the numerical complexity they have to introduce
several approximations in the treatment of stellar lifetimes and yields.

As pointed out by GP\cite{gosw00}, while the prescriptions for the IMF, the
SFR and the infall rate are just empirical recipes due to our poor understanding
of the physical processes involved, the stellar yields can be determined from
first principles, as they rely upon the much better known theory of stellar
evolution. However, there still remain severe uncertainties in the stellar yields,
especially for massive stars where the treatment of convection, the inclusion
or neglect of mass loss, the explosion mechanisms, etc could give rise to large
discrepancies in the final yields of several elements.

Most of the available models use metallicity-independent yields and/or concentrate
on a limited number of chemical species. TWW\cite{timm95} were the first to
consider the evolution of the complete ensemble of light and intermediate-mass
elements (from H to Zn) by including in a simple infall model for the galactic
disk the yields of massive stars with metallicities between Z=0 and Z=Z\( _{\sun } \)
of Woosley \& Weaver (\cite{woos95}), and adopting for the CNO yields of low
and intermediate mass stars the Renzini \& Voli (\cite{renz81}) results. Recently,
GP\cite{gosw00} have reanalyzed the evolution of the elements from C to Zn
by means of an infall model that treats separately the halo and the disk, but
they only take into account stellar yields from massive stars, assuming zero
net yields for intermediate mass stars.

In this paper we present new calculations of the evolution in the solar neighborhood
of all the chemical elements up to Zn , in the framework of a two-infall model
(Chiappini et al. \cite{chia97}) for the formation of the Galaxy, with metallicity
dependent stellar yields for the \textbf{whole range of stellar masses considered}.
Besides, we include the contributions for novae nucleosynthesis, since important
amounts of several isotopes (mainly \element[][7]{Li}, \element[][13]{C}, \element[][15]{N},
and \element[][17]{O}) could be produced by nova outbursts. The plan of the
paper is as follows: In Sec. 2 we present and discuss the main ingredients of
our model: infall assumptions, SFR, IMF and nucleosynthetic prescriptions. Inspired
by the recent observations of Wakker et al. (\cite{wakk99}), who have reported
the detection of a massive cloud with a metallicity \( \sim  \)0.09 times solar
falling into the galactic disk, we consider as our standard model one in which
first the halo and the thick disk of the Milky Way form by accretion of primordial
gas, and then, in a second infall episode lasting up to nowadays, the thin disk
assembles from slightly enriched extragalactic material; however, we have also
considered a more standard model with infall of primordial gas along the whole
galactic evolution. In Sec. 3 we confront our results with the main observational
constraints in the local disk (G-dwarf distribution, age-metallicity relation,
solar abundances, supernova rates, etc.). The core of the paper is presented
in Sec. 4, where we compare in detail the evolution of the different elements
studied with the available observational data, including the most recent observations.
Finally, in Sec. 5 we give the main conclusions of our work.

\section{The chemical evolution model}

We have developed a standard open chemical evolution model where the Milky Way
builds gradually up by infall of primordial and, lately, slightly enriched gas.
No outflows are considered. The disk is divided into concentric independent
rings 1 kpc wide, and we neglect any radial flow between them. Inside each zone
we assume instantaneous mixing so that the stellar ejecta are completely mixed
with the interstellar medium as soon as the stars die; in this way, each ring
is composed by an homogeneous mixture of gas, stars and stellar remnants, and
the local interstellar medium is characterized at any time by a unique composition;
therefore, the quantities describing the state of each zone, i.e. surface density
of total, gas and stellar masses, chemical abundances, etc., are only functions
of galactocentric radius and time. We relaxed the instantaneous recycling approximation
by treating in detail the delay in chemical enrichment due to the finite stellar
lifetimes, which we adopt from the work of the Geneva group (Schaller et al.
\cite{scha92}; Charbonnel et al. \cite{char96}).

The model solves numerically the classical set of nonlinear-integro differential
equations of galactic chemical evolution (Tinsley \cite{tins80}; Pagel \cite{page97}).
In the solar ring we follow the evolution of 76 isotopes from hydrogen to zinc.
To keep handy the computation we use, as in TWW\cite{timm95}, Gaussian quadrature
summation for the mass integrals (Press et al. \cite{press92b}), and a Cash-Karp
stepper method (Press \& Teukolsky \cite{press92a}) for the explicit time integrations.
In the following subsections we discuss the main ingredients of the model.

\subsection{Infall}

Closed box models of galactic evolution face the so called ``G-dwarf problem'',
the formation of too many long lived stars at low metallicities. The most common
way out of this problem is to turn to open models which consider that the Galaxy
forms by continuous infall of extragalactic material. In fact, there are observational
indications of current infall onto the Galactic disk from external regions in
the form of High Velocity Clouds moving towards the Galactic disk, as first
suggested by Larson (\cite{lars72}). Although the interpretation of such clouds
as gas of extragalactic origin has been a matter of debate, the observation
of both High (Mirabel \cite{mira81}) and Very High Velocity Clouds (Mirabel
\& Morras \cite{mira84}) seemed to confirm the idea of current infall with
rates of the order of 1 M\( _{\sun } \) yr\( ^{-1} \). Recently Wakker et
al. (\cite{wakk99}) have reported the detection of a massive (\( \sim  \)10\( ^{7} \)
M\( _{\sun } \) ) cloud falling into the disk of the Milky Way, with a metallicity
0.09 times solar, and the authors give strong arguments for the extragalactic
origin of the cloud.

Different types of infall have been explored in several models, but not all
of them can solve the G-dwarf problem. For instance, constant infall rates just
balancing star formation produce too many stars at high metallicity. Most of
the ``successful'' models use simple exponentially decreasing infall rates
(TWW\cite{timm95}; Prantzos \& Aubert \cite{pran95}; Thomas et al. \cite{thom98})
for the assembling of the galactic disk, with timescales of the order of 3-4
Gyr at the solar ring, but recent determinations of the metallicity distribution
of disk stars by Wyse \& Gilmore (\cite{wyse95}) and Rocha-Pinto \& Maciel
(\cite{roch96}) require, to be reproduced by the models, longer timescales,
with typical values of 7 Gyr at solar galactocentric distances (Chiappini et
al. \cite{chia97}; Prantzos \& Silk \cite{pran98}; Boissier \& Prantzos \cite{bois99}).
It is worth to mention that some chemodynamical models also find long timescales
for the formation of the disk (Samland et al. \cite{saml97}).

Recently Chiappini et al. (\cite{chia97}), and lately Chang et al. (\cite{chan99}),
have included implicitly the halo phase evolution trough models that assume
two subsequent infall episodes. In the first episode the halo and the thick
galactic disk form in a very short time (\( \tau _{T}\leq  \) 1 Gyr). At the
end of this phase, the thin disk begins to form in a second infall episode characterized
by much longer timescales. In this way, the material that eventually forms the
disk has an extragalactic origin.

In the present work, following Chiappini et al. (\cite{chia97}), we adopt the
two-infall exponentially decreasing model. Therefore, the time evolution of
the total surface mass density in the solar ring is given by
\[
\frac{d\sigma (r_{\sun },t)}{dt}=A(r_{\sun })e^{-t/\tau _{T}}+B(r_{\sun })e^{-(t-t_{max})/\tau _{D}}\]
 where \( \tau _{T} \) and \( \tau _{D} \) are respectively the timescales
for the halo-thick disk and thin disk phases, and t\( _{max} \) is the time
of maximum mass accretion onto the thin disk, which corresponds to the end of
the halo-thick disk phase. We set \( \tau _{T} \)=1 Gyr, and the same value
for t\( _{max} \), and we take a value of 7 Gyr for \( \tau _{D} \), the timescale
for the thin disk formation at the position of the Sun (we adopt a solar galactocentric
distance r\( _{\sun } \) = 8.5 kpc).

The coefficients A(r\( _{\sun } \)) and B(r\( _{\sun }) \) are fixed by imposing
that the current total surface mass densities of the thick and thin disks in
the solar neighborhood are well reproduced by the model
\[
A(r_{\sun })=\frac{\sigma _{T}(r_{\sun },t_{G})}{\tau _{T}(1-e^{-t_{G}/\tau _{T}})}\]
 
\[
B(r_{\sun })=\frac{[\sigma (r_{\sun },t_{G})-\sigma _{T}(r_{\sun },t_{G})]}{\tau _{D}[1-e^{-(t_{G}-t_{max})/\tau _{D}}]}\]
 where \( \sigma (r_{\sun },t_{G}) \) and \( \sigma _{T}(r_{\sun },t_{G}) \)
are, respectively, the local surface mass densities of total and thick disk
at the present time, t\( _{G} \), that we take as 13 Gyr. For \( \sigma (r_{\sun },t_{G}) \)
we adopt a value of 54 M\( _{\sun } \) pc\( ^{-2} \) (Rana \cite{rana91};
Sacket \cite{sack97}). The local surface mass density of the thick disk is,
in fact, a parameter in this kind of models. Observational estimates can be
obtained from studies of the density ratio of the thick and thin disks, together
with values of their respective scale heights (Kuijken \& Gilmore \cite{kuij89};
Reid \& Majewski \cite{reid93}; Robin et al. \cite{robi96}; Buser et al. \cite{buse98}).
Unfortunately, the results obtained in those studies span a rather wide range,
from 4.85 to 14.1 M\( _{\sun } \) pc\( ^{-2} \). In our calculations we adopt
a value of \( 10 \) M\( _{\sun } \) pc\( ^{-2} \).

Based on the observations of Wakker et al. (\cite{wakk99}), we begin accreting
primordial material during the halo-thick phase, and then, when the thin disk
initiates its formation, we assume that the infalling material has already being
slightly enriched, with a typical metallicity of 0.1 Z\( _{\sun } \) in solar
proportions. By comparison, we have also calculated models that only accrete
primordial matter during the whole evolution. Even if there are not substantial
differences between the results we obtain in both types of models, in agreement
with Tosi (\cite{tosi88}), who showed that as long as the metallicity of the
accreted material remains below 0.1 Z\( _{\sun } \) there are little changes
in the model results, we obtain slightly better agreement with the observational
constraints in the solar region for the model that incorporates enriched infall
during the assembling of the thin disk.

\subsection{Star formation rate}

In view of the difficulties to understand the rather complicated process of
star formation, standard models for the chemical evolution of the Galaxy adopt
different analytical prescriptions for the star formation rate (SFR) in terms
of intrinsic parameters of spiral galaxies. The simplest and still commonly
used law for star formation is the Schmidt (\cite{schm59}) law: \( \Psi \propto \sigma ^{k}_{g} \),
proportional to some power (between 1 and 2) of the surface gas density. Observations
by Kennicutt (\cite{kenn98}) of the correlation between average SFR and surface
gas densities (total, atomic plus molecular) in spiral and starburst galaxies
point to a value of the exponent in the Schmidt law of \( k \)\( \sim  \)1.5.
Besides this global behaviour, the SFR has to show also dependency on the local
environment, which in turn is typically a function of the galactocentric distance.

The precise radial dependence of the SFR varies according to the scenario considered
for star formation. In theories that describe star formation as a local, self-regulating
process through the balance between the gravitational settling of the gas onto
the disk, that enhances star formation, and the energy injected back into the
interstellar medium by massive young stars under the form of winds and supernova
explosions, which heat and expand the gas, reducing the process of star formation
(Talbot \& Arnett \cite{talb75}), the SFR is a function of the local gravitational
potential and, therefore, of the total surface mass density, \( \sigma  \).
In the original Talbot \& Arnett (\cite{talb75}) formulation, \( \Psi \propto \sigma ^{k-1}\sigma ^{k}_{g} \).
We notice that some chemodynamical models (Burkert et al. \cite{burk92}) find
a similar dependence of the SFR on the total surface mass density. 

Observational support for such star formation law has been obtained by Dopita
\& Ryder (\cite{dopi94}), who showed the existence in spiral disks of an empirical
link between the H\( _{\alpha } \) emission, tracing current star formation,
and the I-band surface brightness, a measure of the contribution of the old
stellar component and, therefore, of the total surface mass density. This observed
relation is well fitted by a SFR law very similar to that of Talbot \& Arnet
(\cite{talb75}): \( \Psi \propto \sigma ^{n}\sigma _{g}^{m} \), with \( n=1/3 \)
and \( m=5/3 \). Moreover, self-regulated star formation can accommodate the
observed correlation between surface brightness and oxygen abundance in late
spiral disks (Edmunds \& Pagel \cite{edmu84}; Ryder \cite{ryde95}). 

We incorporate in our numerical code this law for the star formation as
\[
\Psi (r,t)=\nu \frac{\sigma ^{n}(r,t)\sigma _{g}^{m}(r,t)}{\sigma
^{n+m-1}(r_{\sun },t)}\ \mathrm{M_{\sun } \ pc^{-2} \ Gyr^{-1}}\]
 where the denominator is introduced as a normalization factor in order to express
the efficiency coefficient \( \nu  \) in Gyr\( ^{-1} \). Here we adopt \( \nu = \)1.2
Gyr\( ^{-1} \).

\subsection{Initial mass function}

The initial mass function (IMF) constitutes another basic input in models of
chemical evolution, since it determines in which proportions stars of different
masses enter into play; that, in turn, fixes the averaged stellar yields and
remnants masses for each generation of stars.

In general, the assumed form of the IMF is a declining function of mass in terms
of a power law: \( \Phi (M)\propto M^{-x} \), constant in space and time. Salpeter's
(\cite{salp55}) version, with \( x=1.35 \) for the whole mass range has been
commonly used. However, recent studies reviewed by Scalo (\cite{scal98}), Kroupa
(\cite{krou98}) and Meyer et al. (\cite{meye00}) indicate, within still rather
large uncertainties, that observations are consistent with an IMF almost flat
at low masses, how flat still being a matter of debate (for instance, Reid \&
Gizis (\cite{reid97}) find a unique slope of \( x=0.05 \) for 0.1\( \leq  \)M/M\( _{\sun } \)\( \leq  \)1,
but Kroupa (\cite{krou98}) claims that such slope is only appropriate for M/M\( _{\sun } \)\( \leq  \)0.5),
and declining as a power law with a slope similar to the Salpeter's one above
1 M\( _{\sun } \), although there are again discrepancies on the actual value
of \( x \). For instance, Massey et al. (1995) find slopes \( x \)\( \sim  \)1-1.5
in OB associations, while for massive field stars Massey (\cite{mass98}) obtains
very steep IMFs, with values of \( x \)\( \sim  \)3-4. 

Another aspect of the IMF that has not yet being satisfactorily settled, nor
observationally nor theoretically, is its time behavior. There seems to be a
tendency between observational researchers to favor no variations of the IMF,
although in a recent review Scalo (\cite{scal98}) argues against its universality
in space and time. Besides, an IMF producing more massive stars in the early
Galaxy has been invoked as a solution for the G-dwarf problem. Nevertheless,
in a recent paper, Chiappini et al. (\cite{chia00}) have investigated the effects
on galactic chemical evolution of several time dependent IMFs and conclude that
the combination of infall and a constant IMF is still the best choice to reproduce
the observational constraints in the Milky Way.

In this paper we adopt the constant IMF version of Kroupa et al. (\cite{krou93})
which consists of a three slope power law. In the range of very low masses the
slope is quite flat, \( x \)=0.3 for M\textless{}0.5 M\( _{\sun } \), it steepens
to \( x \)=1.2 in the interval 0.5 M\( \leq  \)1 M\( _{\sun } \), and in
the high mass regime M\textgreater{}1 M\( _{\sun } \) the slope agrees with
the one by Scalo (\cite{scal86}) \( x \)=1.7. As usual, we normalize to unity
this IMF between a minimum stellar mass of 0.08 M\( _{\sun } \), the H-burning
limit, and a maximum of 100 M\( _{\sun } \).

\subsection{Nucleosynthesis prescriptions}

The chemical yields synthesized by stars in different stellar mass ranges are
a key ingredient when trying to understand the chemical evolution of galaxies,
in particular those corresponding to ``massive stars'', stars that end their
lives in Type II supernova explosions, since they are the main responsibles
for the enrichment of the Universe in intermediate and heavy nuclei, through
both hydrostatic burning (elements up to calcium) and explosive nucleosynthesis
(iron peak elements, to which Type Ia supernovae are also important contributors).
Even if the uncertainties in the theory of stellar evolution are far less important
than those affecting the SFR or the IMF, there still remain serious discrepancies
in the literature about the treatment of crucial aspects of stellar evolution.
For instance, presupernova configurations are affected by the precise formulation
of convection, semi-convection and overshooting, the adopted nuclear reaction
rates (especially for the \element[][12]{C}(\( \alpha ,\gamma  \))\element[][16]{O}
reaction), the inclusion or not of mass loss and rotation, etc. A similar situation
holds for the results of core collapse supernova explosions. In fact, in current
calculations the explosion itself is induced rather arbitrarily and the explosion
energy is fixed by imposing a given value for the final kinetic energy. Also,
the ejected mass crucially depends on the degree of ``fall-back'', and thus
on the precise details of the explosion. We thus see that the published stellar
yields are yet plagued with severe uncertainties.

Although the evolution during the thin disk epoch will be hardly affected by
the slight metallicity dependence shown in the published stellar yields, the
early halo-thick disk phase could be strongly influenced by the material ejected
by stars of different metallicities. Hence, we will consider metallicity dependent
yields for the whole stellar mass range.

\subsubsection{Massive stars: 8 M\protect\( _{\sun }\leq \protect \)M\protect\( \leq \protect \)100
M\protect\( _{\sun }\protect \)}

We assume that stars in this mass interval end their lifes as core collapse
supernovae. The two major sources in this mass range, widely used in chemical
evolution models, are those of Woosley \& Weaver (\cite{woos95}) (hereafter
WW\cite{woos95}), that calculated full stellar models without mass loss nor
rotation, for stars with masses comprised between 12 M\( _{\sun } \) and 40
M\( _{\sun } \) and with different initial metallicities (Z/Z\( _{\sun } \)=0,
10\( ^{-4} \), 10\( ^{-2} \), 10\( ^{-1} \) and 1), and those of Thielemann
et al. (\cite{thie96}), who calculated the evolution of He cores corresponding
to stars of solar initial metallicity up to 70 M\( _{\sun } \). There are noticeable
differences between the WW\cite{woos95} results for Z=0 and Z\( \neq  \)0,
while the yields and stellar remnant masses for nonzero initial metallicities
give more similar results. Solar composition models of WW\cite{woos95} and
Thielemann et al. (\cite{thie96}) give almost the same values for the yields
of the CNO isotopes, but non-negligible differences exist for other important
elements. For instance, the magnesium yield is systematically lower in WW\cite{woos95}
as a consequence of the different treatments of convection. In the case of the
iron yield, WW\cite{woos95} gives much higher values than Thielemann et al.
(\cite{thie96}) for stars with masses below \( \sim  \)35 M\( _{\sun } \),
since it is very sensitive to the mass cut that separates the ejecta and the
material that falls back onto the compact remnant which, as mentioned above,
depends in turn on the detailed explosion mechanism (see Thomas et al. \cite{thom98},
and Chiappini et al. \cite{chia99} for a complete discussion).

Lately, Limongi et al. (\cite{limo00}), from full stellar models, again without
mass loss nor rotation, have also presented yields of massive stars in the range
13-25 M\( _{\sun } \) for three metallicities (Z=0, 10\( ^{-3} \), and 0.02).
As shown by GP\cite{gosw00}, the main difference with respect to WW\cite{woos95}
is a more marked odd-even effect in the case of Limongi et al. (\cite{limo00})
yields, which translates into systematically lower yields of odd Z elements.

In this paper we adopt the metallicity-dependent yields of WW\cite{woos95}
for massive stars between 8 and 100 M\( _{\sun } \). The reasons for this choice
are twofold: they are the results of full evolutionary calculations and they
consider a wider interval of masses and metallicities as compared with the work
of Limongi et al. (\cite{limo00}). In particular, we take their models A up
to 25 M\( _{\sun } \), and their models B for 30, 35 and 40 M\( _{\sun } \)
(the explosion energy in models B is higher than in models A). Those yields
have been extrapolated up to 100 M\( _{\sun } \), even though this has a minor
effect on the final results since the Kroupa et al. (\cite{krou93}) IMF produces
very few stars more massive than 40 M\( _{\sun } \). 

TWW\cite{timm95} suggested that a better agreement with most of the observed
evolution of the abundances is obtained if the WW\cite{woos95} iron yields
are reduced by a factor of two. Observational estimates of the iron synthesized
in SN 1987A (the explosion of a 20 M\( _{\sun } \) belonging to a stellar system,
LMC, with an estimated Z\( \sim  \)0.1 Z\( _{\sun } \)) and of SN 1993J (whose
progenitor was a 14 M\( _{\sun } \) star in the galaxy M81, where Z\( \sim  \)Z\( _{\sun } \))
also points to such an overestimate (Thomas et al \cite{thom98}). Therefore,
in view of the uncertainties arising from the flaws in the explosion calculations,
we reduce by a factor 2 the nominal yields of WW\cite{woos95}.

\subsubsection{Low and Intermediate Mass Stars: M\protect\( \leq \protect \)8 M\protect\( _{\sun }\protect \)}

As it is well known, single stars in this mass range pollute the interstellar
medium through moderate stellar winds and planetary nebula ejection, ending
their lifes as white dwarfs. These stars make important contributions to the
He, C and N galactic contents. Here again important uncertainties on the final
yields exist that result from the treatment of mass loss, convection, evolution
on the asymptotic giant branch, etc. 

The yields of Renzini \& Voli (\cite{renz81}) have been extensively used in
chemical evolution models. However, in the last years several groups have published
new evolutionary calculations for stars belonging to this mass interval and
for different initial metallicities (Marigo et al. \cite{mari96}; van den Hoek
\& Groenewegen \cite{vand97}). In this work the contribution to the galactic
enrichment by these stars has been taken from van den Hoek \& Groenewegen (\cite{vand97}),
who give yields for stars with masses comprised between 0.9 and 8 M\( _{\sun } \)
and consider five initial metallicities, since the work of Marigo et al. (\cite{mari96})
is limited to stars with M\( \leq  \)4 M\( _{\sun } \), besides of taking
into account convective overshooting, which is not considered in WW\cite{woos95}.

\subsubsection{Type Ia Supernovae}

There is a general agreement that Type Ia supernovae are important contributors
to iron and other iron peak elements (in particular \element[][58]{Ni} and \element[][54]{Cr})
in the late disk evolution, up to the point that they are responsible for \( \sim  \)2/3
of the total iron contents (TWW\cite{timm95}). We adopt the conventional view
that Type Ia supernovae are the carbon deflagration of massive C-O white dwarfs
in binary systems. The contribution of Type Ia supernovae to the chemical enrichment
of the galaxy, specially the iron content, has been calculated following the
prescriptions of Matteucci \& Greggio (\cite{matt86}), where Type Ia supernovae
come from binary systems with a minimum mass of 3 M\( _{\sun } \), in order
to ensure that the accreting white dwarf eventually reaches the Chandrasekhar
mass, and a maximum mass of 16 M\( _{\sun } \), if we assume that C-O white
dwarfs come from primaries up to 8 M\( _{\sun } \). Thus, the amount of isotope
\textit{i} due to this source is given by

\[
C\int ^{16}_{3}\Phi (M_{B})\left[ \int ^{0.5}_{\mu _{m}}f(\mu )\Psi (r,t-\tau _{\mu \cdot M_{B}})\frac{Y_{i}}{M_{B}}d\mu \right] dM_{B}\]

The parameter \emph{C}, which determines the fraction of suitable binary systems
that actually produce Type Ia SN, is set equal to 0.04 in order to fulfill the
restrictions imposed by observational data on supernovae frequency. The chemical
composition of the ejecta are taken from Thielemann et al. (\cite{thie93})
for the well known model W7 (Nomoto et al. \cite{nomo84}), and we also include
the contribution to the enrichment of the interstellar gas by the secondary
star.

\subsubsection{Classical Novae}

Although classical novae have little or no influence on the evolution of the
abundances of most of the chemical elements included in our calculations, according
to José \& Hernanz (\cite{jose98}) they can be significant contributors to
certain isotopes that are strongly overproduced in nova explosions relative
to the solar system abundances (in particular those with overproduction factors
larger than 1000). \element[][7]{Li}, \element[][13]{C}, \element[][15]{N} and
\element[][17]{O} are some of the isotopes whose abundances could be affected
by novae. 

In order to obtain the nova rates we assume that novae take place in binary
systems made of a white dwarf coming from stars with main sequence masses in
the interval \( M_{1}^{m}= \)1 M\( _{\sun } \) to \( M_{1}^{M}= \)8 M\( _{\sun } \),
while the secondary stars have masses between \( M_{2}^{m}= \)0.5 M\( _{\sun } \)
and \( M_{2}^{M}= \)1.5 M\( _{\sun } \). The rate of those explosions can
be estimated by a procedure similar to that used by Matteucci \& Greggio (\cite{matt86})
to calculate the rate of SN Ia:

\begin{eqnarray*} \frac{dR_{outburts}}{dt} & = & D\int_{M(t)+M_{2}^{m}}^{M_{1}^{M}+M_{2}^{M}}\frac{\phi (M_{B})}{M_{B}}dM_{B}\\  & & \int _{\mu _{m}}^{\mu _{M}}f(\mu )\psi (t-\tau _{M_{B}(1-\mu )}-t_{cool})d\mu \end{eqnarray*}

where \( M_{B} \) is the total mass of the system and

\( \mu_M= \left\{ \begin{array}{cl} \frac{M_{2}^{M}}{M_{B}} & M_{B}>2 M_{2}^{M}\\ 0.5 & M_{B}<2M_{2}^{M} \end{array}\right. \)

\( \mu_m= \left\{ \begin{array}{cl} \max \left\{ \frac{M_{B}-M_{1}^{M}}{M_{B}},\frac{M_{2}^{m}}{M_{B}}\right\} & M_{B}>M_{1}^{M}\\ \frac{M_{2}^{m}}{M_{B}} & M_{1}^{M}>M_{B}>2M_{2}^{m}\\ 0.5 & M_{2}^{m}>M_{B} \end{array}\right. \)

In this expression, the value of the free parameter \emph{D}, which plays a
role similar to the parameter \emph{C} in the rate of Type Ia supernovae, is
obtained by imposing that the formula reproduces the observed outburst rate
at the present age of the Galaxy, of the order of \( \sim  \)40 yr\( ^{-1} \)(Hatano
et al. \cite{hata97}), and t\( _{cool} \), the cooling time for a white dwarf
before it can produce the first nova outburst, is taken as 1 Gyr . 

The yields for the classical nova outburst are an average of those given by
José \& Hernanz (\cite{jose98}). We have considered that 30\% of the nova outbursts
come from ONe white dwarf progenitors, while 70\% come from CO ones.

\section{Model results versus observational constraints}

As we have discussed in the previous section, there are large uncertainties
in the ingredients of standard chemical evolution models which translate into
a series of more or less free parameters, e.g. infall timescales, efficiency
and exponent of the SFR, slope of the IMF, etc. Fortunately, for the solar neighborhood,
the number of observational constraints is large enough to restrict the space
of variation of these parameters (Boissier \& Prantzos \cite{bois99}). Results
for the whole galactic disk will be the subject of another paper.

\subsection{Surface densities, SFR and supernova rates}

\begin{figure}
{\par\centering \resizebox*{\hsize}{!}{\includegraphics{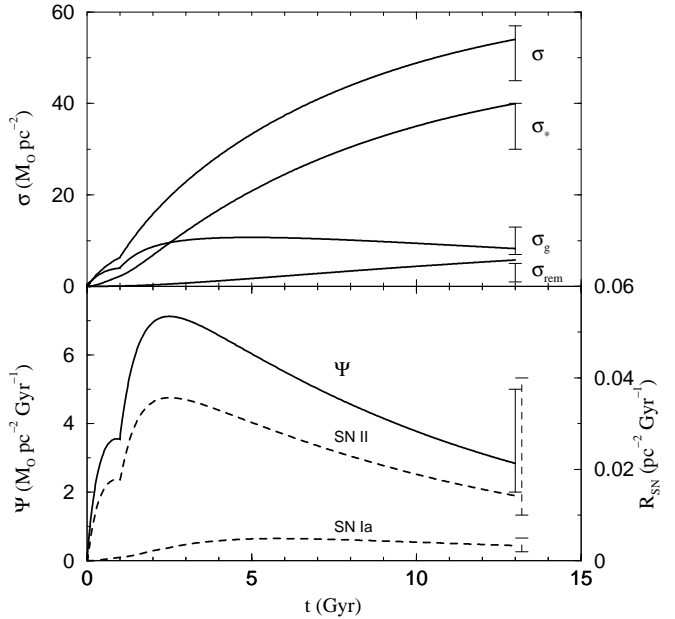}} \par}

\caption{\textit{Upper panel}: time evolution of the surface densities
of total mass (\protect\( \sigma \protect \)), visible stars (\protect\( \sigma _{\star }\protect \)),
gas (\protect\( \sigma _{g}\protect \)), and stellar remnants (\protect\( \sigma _{rem}\protect \))
for our two-infall model. \textit{Lower panel}: evolution of the stellar formation
rate, as well as the Type II and Type Ia supernova rates.}
\label{sigsfr}
\end{figure}

First of all, any successful model has to reproduce the present-day surface
density of total mass, gas, star and stellar remnants, the current rate of star
formation, infall and supernovae (both Type Ia and Type II). In Table \ref{XXtable}
we show our results for the present values (t\( _{G} \)=13 Gyr) of those quantities
compared with the current observational values for the solar vicinity. The time
evolution of the surface densities of total mass (\( \sigma  \)), stars (\( \sigma _{\star } \)),
gas (\( \sigma _{g} \)), and stellar remnants (\( \sigma _{rem} \)) are displayed
in the upper panel of Fig. \ref{sigsfr}, while in the lower panel appear the
star formation rate, and the Type II and Type Ia supernova rates. We see from
Table \ref{XXtable} and Fig. \ref{sigsfr} that the model presented in the
previous section nicely fits the observations.

\begin{table}

\caption{Calculated and observed present values for several relevant
quantities in the solar ring.}
\label{XXtable}
\vspace{0.3cm}
{\centering \begin{tabular}{p{4.5cm}ccc}
\hline 
&
model &
Obser.&
Ref.\\
\hline 
\hline 
\( \sigma _{g} \) (M\( _{\sun } \) pc\( ^{-2} \))&
8.27&
7-13&
1\\
\( \sigma _{\star } \) (M\( _{\sun } \) pc\( ^{-2} \))&
39.93&
30-40&
2\\
\( \sigma _{rem} \) (M\( _{\sun } \) pc\( ^{-2} \))&
5.77&
2-4&
3\\
\( \sigma _{g}/\sigma _{tot} \)&
0.15&
0.05-0.2&
\\
\( \sigma _{inf} \) (M\( _{\sun } \) pc\( ^{-2} \) Gyr\( ^{-1} \))&
1.38&
0.3-1.5&
4\\
\( \Psi  \) (M\( _{\sun } \) pc\( ^{-2} \) Gyr\( ^{-1} \))&
2.84&
2-5&
5\\
SN Ia per century&
0.41&
0.17-0.7&
6\\
SN II per century&
1.76&
0.55-2&
6\\
SN II/SN Ia&
4.3&
3-6&
6\\
Novae (yr\( ^{-1} \))&
40.0&
20-60&
7\\
\hline 
\end{tabular}\par}\vspace{0.3cm}

{\small REFERENCES.-(1) Dickey \cite{dick93}; Flynn et al. \cite{flyn99};
(2) Gilmore et al. \cite{gilm89}; (3) Méra et al. \cite{mera98}; (4) Portinari
et al. \cite{port98} and references therein; (5) Güsten \& Mezger \cite{gust82};
(6) Tamman et al. \cite{tamm94}; (7) Hatano et al. \cite{hata97}}{\small \par}
\end{table}

\subsection{Age-metallicity relation}

Our model also fulfills the requisite of producing an age-metallicity relation
in agreement with the observed one. The age-metallicity relation (AMR) shows
the evolution of the ratio {[}Fe/H{]}, taken as a measure of the metallicity
of the Galaxy, as a function of time. The pioneering work by Twarog (\cite{twar80})
showed that {[}Fe/H{]} rises steeply during the first 2 Gyr up to a value of
\( \sim  \)\( - \)0.5 dex, and afterwards increases smoothly with time towards
the solar value, although the data showed a large dispersion, and age-bins and
average metallicity per bin were used. Later reexamination of Twarog's data
by Meussinger et al (\cite{meus91}) found a similar AMR. More recently, Edvardsson
et al. (\cite{edva93}), based on a different sample of nearby F and G stars,
obtained for binned data a good agreement with previous results, but they claim
that the large dispersion found in their unbinned data was essentially real
and not observational (but see Garnett \& Kobulnicky \cite{garn00}). Finally,
Rocha-Pinto et al. (\cite{roch00}) obtained a chromospheric AMR using a sample
of 552 late-type dwarfs that basically confirms the observed trends, but with
almost half the scatter found by Edvardsson et al. (\cite{edva93}).

Models of chemical evolution that adopt the instantaneous mixing approximation,
like ours, cannot produce any dispersion at all, and at most they can only fit
the mean relation. Therefore, the AMR is a less tight constraint, even though
the average relation must be fitted by successful models.

\begin{figure}
{\par\centering \resizebox*{\hsize}{!}{\includegraphics{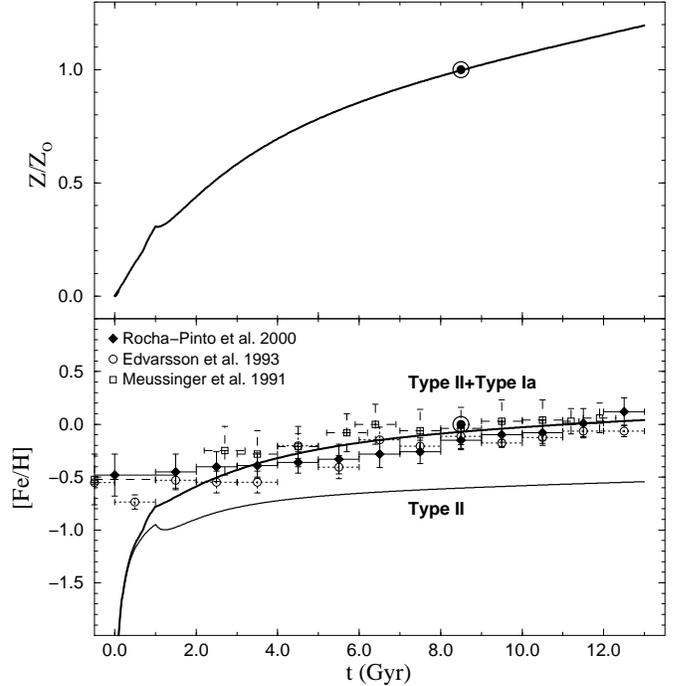}} \par}

\caption{\textit{Upper panel}: evolution with time of the total metallicity
Z normalized to the solar value Z\protect\( _{\sun }\protect \). \textit{Lower
panel}: time evolution of the ratio {[}Fe/H{]} for contributions from Type II
supernovae (\textit{thin line}) and for Type II plus Type Ia supernovae (\textit{thick
line}). }
\label{AMRZ}
\end{figure}

We present in the lower panel of Fig. \ref{AMRZ} our results for the evolution
with time of {[}Fe/H{]} compared with the AMR by Meussinger et al. (\cite{meus91}),
Edvarsson et al. (\cite{edva93}) and with the more recent data of Rocha-Pinto
et al. (\cite{roch00}). Just for completeness, in the upper panel we also show
the evolution of the total metallicity in terms of the solar value. Again our
model fits well the averaged trend of the data, although, as stated above, this
is not a really tight constraint, because almost any model can produce a reasonable
AMR.

Nearly 2/3 of the present total iron contents are made by Type Ia supernovae,
the other third coming from Type II supernovae. We recall here that the iron
yields for Type II supernovae used in this work are half those of WW\cite{woos95}
in order to obtain a better fit to the evolution of the element abundances.
That is also indicated by iron abundance measurements in supernova ejecta. In
any case, as pointed out by TWW\cite{timm95}, whatever contribution to the
present galactic iron contents by Type II supernovae comprised between one and
two thirds is allowed by our current understanding of stellar physics.

\subsection{G-dwarf distribution}

\begin{figure*}
{\par\centering \resizebox*{\hsize}{!}{\includegraphics{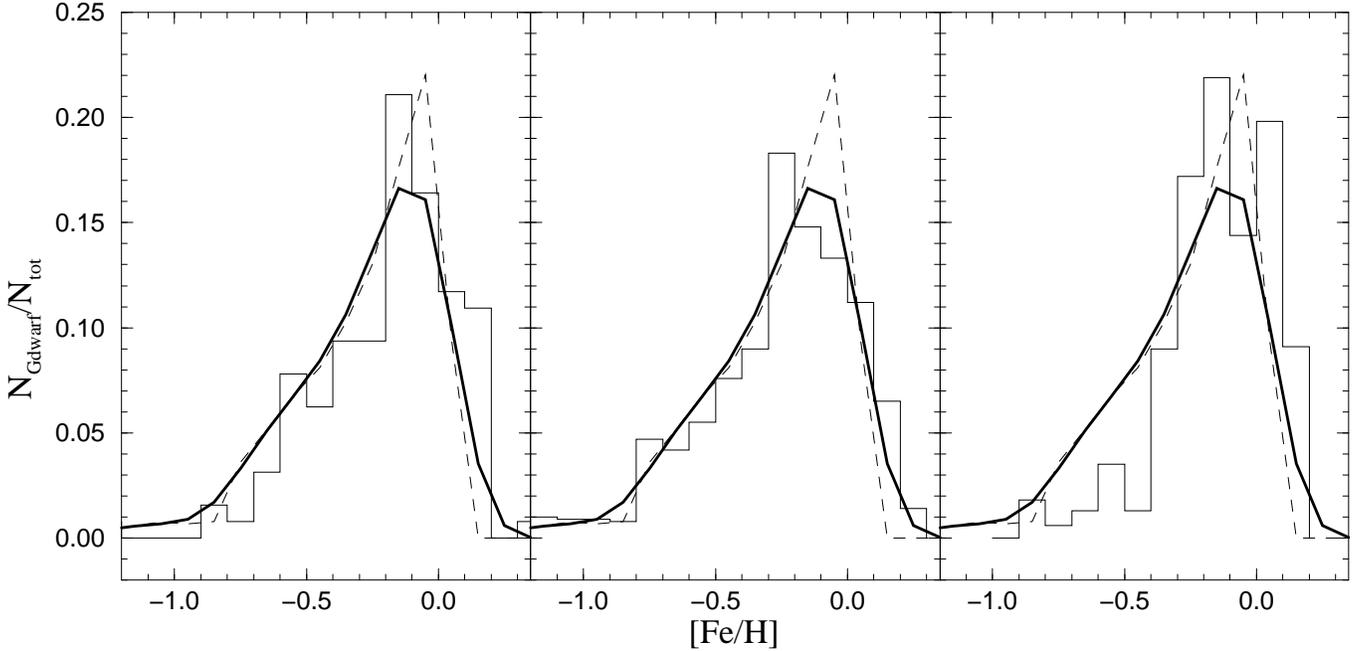}} \par}

\caption{Rough (\textit{thin dashed curve}) G-dwarf metallicity distributions
and Gaussian convolved distribution (\emph{thick solid curve}) compared with
observational data by Wyse \& Gilmore (\cite{wyse95}) (\textit{left panel}),
Rocha-Pinto \& Maciel (\cite{roch96}) (\textit{central panel}) and by J\o rgensen
(\cite{jorg00}) (\textit{right panel}).}
\label{Gdwarf}
\end{figure*}

The stars of spectral type G have main-sequence lifetimes comparable or even
larger than the estimated age of the Galaxy. Hence, the distribution in metallicity
of a complete sample of these stars in the solar neighborhood carries memory
of the star formation history. This makes the observed G-dwarf distribution
one of the more stringent constraints for models of galactic chemical evolution. 

As it is well known, the paucity of G-dwarf stars at low metallicities, referred
to as the ``G-dwarf problem'', cannot be explained by simple, closed box models.
Several solutions have been proposed to solve this problem, but open models
with progressive infall of primordial or slightly enriched material with long
timescales for the disk formation, like ours, are still the best option (Chiappini
et al. \cite{chia97}), besides of being compatible with dynamical simulations
of the formation of galactic disks (Burkert et al. \cite{burk92}), and with
the observation of infall of High and Very High Velocity Clouds.

Fig. \ref{Gdwarf} compares the predicted G-dwarf distribution produced by our
model with the most recent observational data from Wyse \& Gilmore (\cite{wyse95})
(\textit{left panel}), Rocha-Pinto \& Maciel (\cite{roch96}) (\textit{central
panel}) and J\o rgensen (\cite{jorg00}) (\textit{right panel}). The direct
results are displayed by thin dashed lines, while the thick solid ones show
the convolution with a Gaussian with a dispersion of 0.15 in order to simulate
observational and intrinsic scatter. Very good agreement is reached, specially
for the rising part of the distribution and the location and height of the peak
when comparing with the two first sets of data. In the case of the distribution
from J\o rgensen (\cite{jorg00}), the fit is still reasonable, the peak is
again well reproduced, but the model shows a low metallicity tail (around {[}Fe/H{]}\( \sim  \)
\( - \)0.5) that is higher than the data. The comparison between theoretical
results and observations could be improved by relaxing the instantaneous mixing
approximation, introducing thus in the models the intrinsic scatter. Our results
confirm once more that infall models with long timescales for the assembling
of the disk are capable of solving the G-dwarf problem. 

\begin{figure}
{\par\centering \resizebox*{\hsize}{!}{\includegraphics{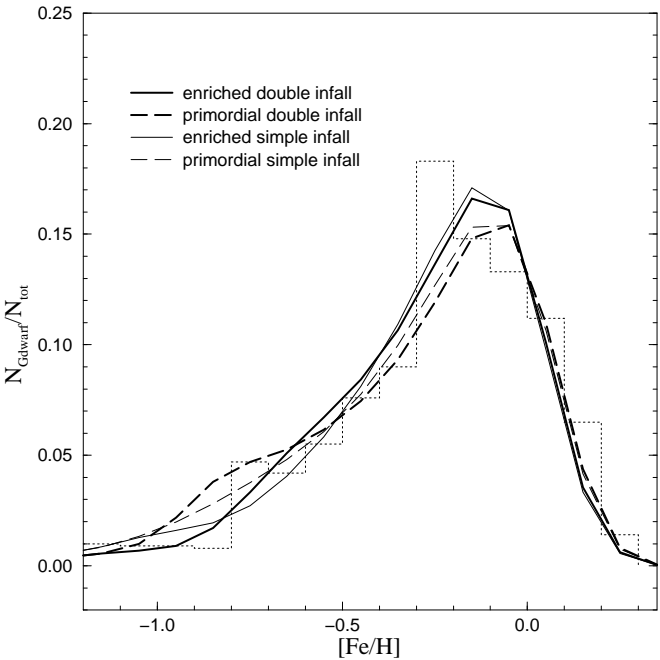}} \par}

\caption{G-dwarf distributions obtained for different types of infall.
The observational data correspond to Rocha-Pinto and Maciel (\cite{roch96}).}
\label{Gdwarfvar}
\end{figure}

As indicated above, our model uses a double exponential infall in order to treat
as separate entities the halo-thick disk and the thin disk, and we consider
accretion of enriched material (Z=0.1Z\( _{\sun } \)) during the thin disk
phase. We have also calculated models that incorporate primordial material both
in the halo-thick disk and in the thin disk phases, as well as models that only
consider simple exponential infall of primordial and primordial plus enriched
material which, if not appropriate for the halo and thick disk, are relevant
when making comparisons for the thin disk properties. In all those models we
have adopted the same timescale for the galactic disk, i.e., 7 Gyr. We obtain
in all cases very similar results for the final characteristics of the solar
ring, but slight differences appear for the convolved G-dwarf distributions,
which are displayed in Fig. \ref{Gdwarfvar}, where for clarity only the observations
by Rocha-Pinto \& Maciel (\cite{roch96}) are shown. One and two infall models
with the same composition of the incorporated matter are almost indistingishable,
but primordial composition models show a moderate excess of stars in the low
metallicity tail and lower maxima than those obtained in enriched models.

\subsection{Solar abundances}
\begin{figure*}
{\par\centering \resizebox*{\hsize}{!}{\includegraphics{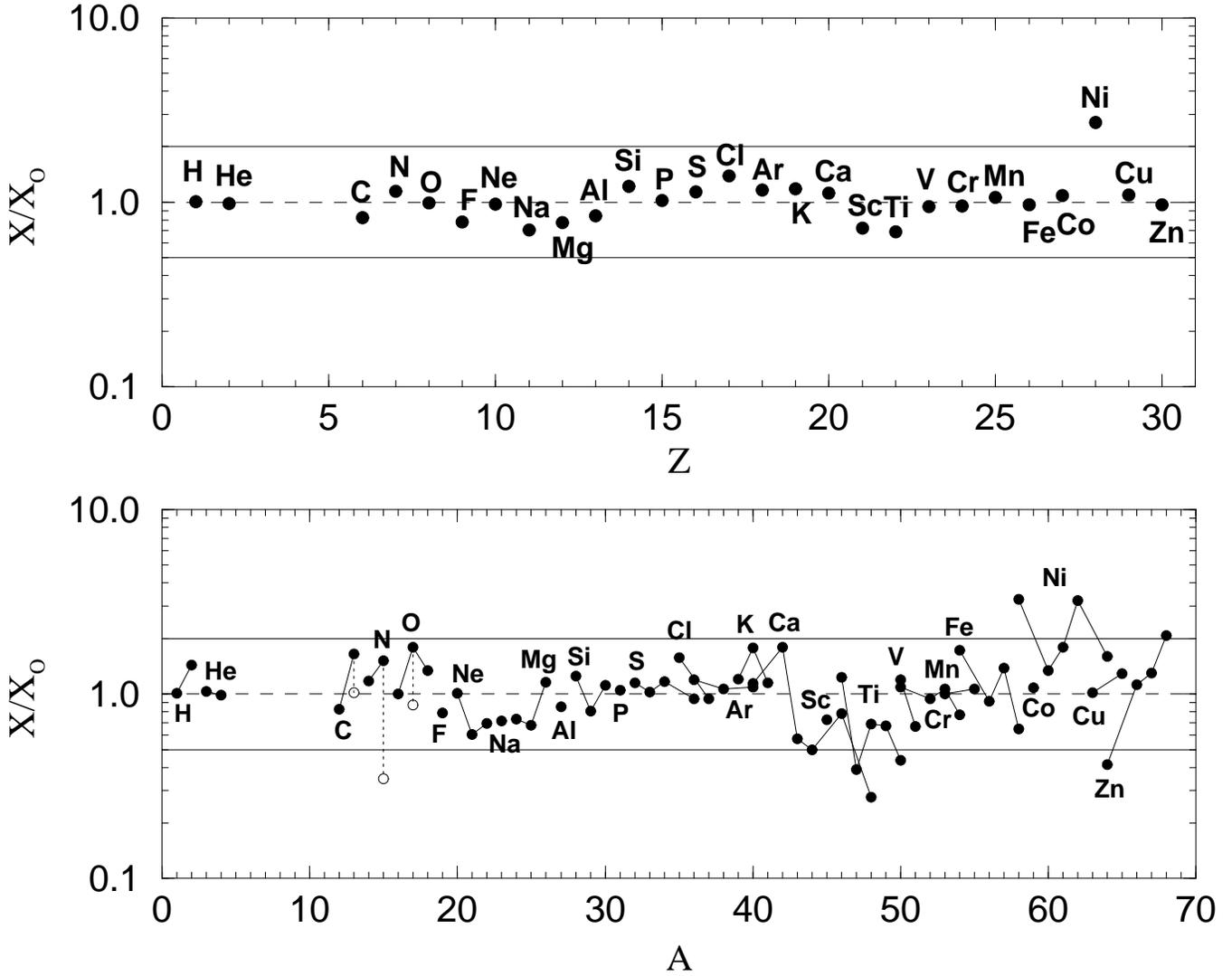}} \par}

\caption{Mass fractions of the calculated elements (\textit{upper panel}),
from hydrogen to zinc (lithium, beryllium and boron are not shown), and of all
their stables isotopes (l\textit{ower panel}) in the interstellar gas at the
time of solar birth, relative to the solar abundances of Anders \& Grevesse
(\cite{ande89}). The isotopes of the same element are connected by solid lines.
The dashed horizontal line would correspond to a perfect agreement, while the
two solid lines correspond to a discrepancy by a factor of 2. We take as successful
those elements and isotopes whose calculated abundance fall within this factor
of 2 of their solar value. The three open circles in the lower panel represent
the relative abundances of \element[][13]{C}, \element[][15]{N} and \element[][17]{O}
when nova yields are excluded from the calculations.}
\label{XXsun}
\end{figure*}

We present in Fig. \ref{XXsun} the calculated mass fractions of the elements
(upper panel) and all their stable isotopes (lower panel) studied in this work
(except LiBeB isotopes, whose abundances have important contributions from galactic
cosmic ray nucleosynthesis, not included in the present calculations) in the
interstellar gas at the time of the solar birth, 4.5 Gyr ago, compared with
the observed solar values (Anders \& Grevesse \cite{ande89}), considered as
representative of the composition of the interstellar gas at that epoch. We
must recall here that this last assumption is far from being definitely settled,
since measured abundances in regions of recent star formation, like the Orion
nebula, show metallicities lower than solar (see, for instance, Cardelli \&
Federmann \cite{card97}).
 
Due to the uncertainties existing in the observational values, we consider as
good agreement abundances within a factor of 2 of their solar value. As it can
be seen in Fig. \ref{XXsun}, all the elemental abundances fulfill this criterion,
nickel excepted, which is clearly oversolar, and the model makes the correct
solar metallicity. The isotopic abundances are also nicely adjusted. The spread
for isotopes below calcium is lower than above it, which reflects the uncertainties
in the present modeling of Type II supernova explosions.

The production factors for hydrogen, helium and the most important CNO isotopes
are nearly equal to unity. In particular, the solar \element[][3]{He} abundance
is almost exactly reproduced by our models without taking into account any partial
destruction by extra-mixing on the RGB (see, however, Romano et al. \cite{roma00}).
The inclusion of novae worsens our results for \element[][13]{C} and \element[][17]{O},
which are then a little bit overabundant. On its turn, \element[][15]{N} is
a factor of 3 below its solar value when only nucleosynthesis in massive stars
is considered, and contribution by novae is mandatory to render its abundance
compatible with the solar abundance.

A few isotopes fall out of the acceptable range. \element[][48]{Ca}, \element[][47]{Ti},
\element[][50]{Ti} and \element[][64]{Zn}, are underproduced (especially \element[][48]{Ca}).
In the case of the neutron-rich isotopes \element[][48]{Ca} and \element[][50]{Ti}
there could be a substantial contribution from Type Ia supernovae resulting
from initially slow deflagrations at high density (Woosley \& Eastman \cite{woos94a}).
\element[][47]{Ti} (and, in lesser extent, \element[][44]{Ca} and \element[][51]{V})
have been traditionally problematic; these isotopes could have noticeable contributions
from sub-Chandrasekhar models of SNIa (Shigeyama et al. \cite{shig92}; Woosley
\& Weaver \cite{woos94b}). Additional contributions to the abundance of \element[][64]{Zn}
could be obtained, as pointed out by TWW\cite{timm95}, from weak s-process
in massive stars or classical s-process in low-mass AGB stars.

On the other hand, there is a small overproduction of nickel, due to the large
yields of \element[][58]{Ni}\( ^{58} \)Ni in the adopted model for SNIa and
the high production of \element[][62]{Ni}\( ^{62} \)Ni in current models of
massive stars explosions which reflects the inaccuracies involved in present
supernova modeling. 

As noted in previous works (TWW\cite{timm95}; GP\cite{gosw00}), the fact that
the solar abundances of all isotopes up to the iron peak, that cover a range
of almost 9 orders of magnitude, are so precisely reproduced by current models
of chemical evolution indicates that the present understanding of stellar nucleosynthesis
is basically correct, at least to first order.

\section{Abundance ratios evolution}

\begin{figure}
{\par\centering \resizebox*{\hsize}{!}{\includegraphics{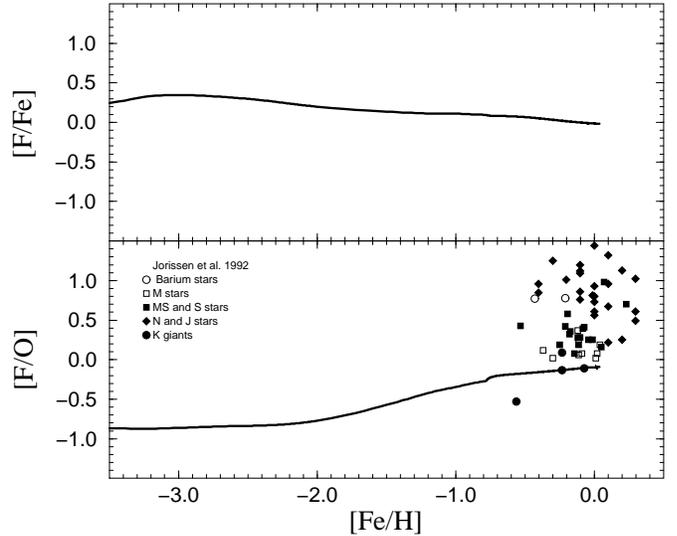}} \par}

\caption{\textit{Upper panel}: evolution of the ratio {[}F/Fe{]} as a
function of {[}Fe/H{]}. \textit{Lower} \textit{panel}: evolution of the ratio
{[}F/O{]} as a function of {[}Fe/H{]}, compared with observational points from
Jorissen et al. (\cite{jori92}).}
\label{abunF}
\end{figure}

We have calculated the evolution in time of all the stable isotopes of thirty
chemical elements from H to Zn with the nucleosynthetic prescriptions detailed
above. Our results are displayed in Figs. \ref{abunCNO} to \ref{abunCuZn},
where we compare the predicted ratios {[}X/Fe{]} as functions of {[}Fe/H{]}
with observational data of spectroscopic abundances in nearby stars. Data sources
are detailed in the legends of the figures. As the data normally refer to elemental
abundances, for each element we sum over the calculated isotopic abundances.

Before going on with the presentation of our results we wish to recall again
that following the suggestion of TWW\cite{timm95} we have reduce the iron yields
of WW\cite{woos95} by a factor of 2. The reduced iron yields, well within the
current inaccuracies in supernova models, not only give a better agreement with
the observed elemental abundances but are also consistent with the amount of
iron ejected in recently observed Type II supernovae .

We consider that the calculated abundance evolution is only significant for
{[}Fe/H{]}\( \geq  \)\( - \)3, a value that our model reaches at t\( \sim  \)30
Myr, which approximately corresponds to the lifetime of a 8-9 M\( _{\sun } \)
star. This means that only after reaching such global metallicity the yields
of massive stars are fully averaged over the IMF, therefore reducing the possible
uncertainties in the yields of individual stars.

\subsection{The CNO elements}

\begin{figure}
{\par\centering \resizebox*{\hsize}{!}{\includegraphics{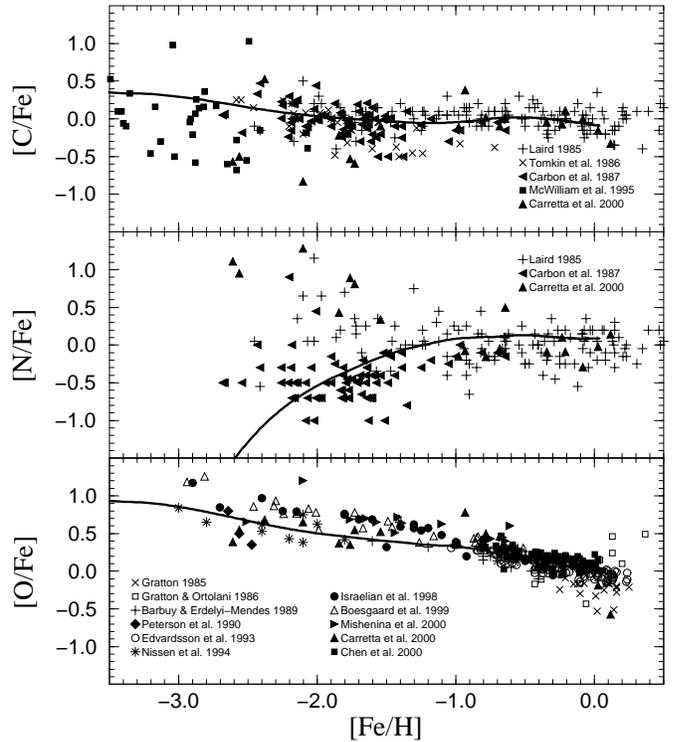}} \par}

\caption{Evolution of abundances ratios {[}X/Fe{]} as a function of
{[}Fe/H{]} for the CNO elements. \textit{Upper panel}: Carbon; \textit{central
panel}: Nitrogen; l\textit{ower pane}l: Oxygen}
\label{abunCNO}
\end{figure}

We present in Fig. \ref{abunCNO} the results obtained in our calculations for
the evolution of CNO elemental abundances with respect to iron.

Carbon and iron are primary elements and are produced by very different processes:
carbon is made by the triple-\( \alpha  \) process in hydrostatic helium burning
in massive and, mostly, in intermediate and low-mass stars, while Fe is produced
by explosive burning in Type II and Type Ia supernovae. Therefore, we expect
that the ratio {[}C/Fe{]} remains nearly constant as a function of {[}Fe/H{]}.
Observations of carbon abundances in halo and disk dwarfs (field giants are
not reliable indicators of ab initio carbon abundance due to the effects of
the first dredge-up) do actually show that {[}C/Fe{]} is almost constant in
time, with a value approximately solar (Laird \cite{lair85}; Tomkin et al.
\cite{tomk86}; Carbon et al. \cite{carb87}; McWilliam et al. \cite{mcwi95}),
although there is an important dispersion in the data at all metallicities.
Those surveys also seem to indicate some trend towards slightly higher values
of {[}C/Fe{]} at low {[}Fe/H{]}, as confirmed to first order by Wheeler et al.
(\cite{whee89}) after reanalyzing the data available at that epoch, although
observational uncertainties counsel to take with some caution this upturn at
low metallicity.

Our results show that the synthesis of carbon and iron by massive stars suffices
to explain the observations in halo stars. The {[}C/Fe{]} ratio shows a slow
decline towards the solar value during the halo-thick disk epoch, which indicates
that intermediate-mass stars, whose lifetimes permit them to evolve during the
late halo phase, do not give very important contributions to the carbon contents.
The bump around {[}Fe/H{]}\( \sim  \)\( - \)1, at the beginning of the thin
disk phase, results from the contribution of intermediate and, specially, low-mass
stars (M \textless{} 2 M\( _{\sun } \)), which eject important amounts of carbon
but no iron. But the iron ejected by Type Ia supernovae, which start to contribute
around the same epoch, compensates and finally overwhelms the role of the low
mass stars.

The synthesis of \element[][13]{C} in intermediate and low-mass stars is capable
of building up the solar contents. The inclusion of novae produces a small overabundance
of \element[][13]{C} at the time of the solar birth.

The evolution of the ratio {[}N/Fe{]} as a function of {[}Fe/H{]} is displayed
in the second graph of Fig. \ref{abunCNO}. The three observational surveys
shown in the figure present a large scatter and there are no observations below
{[}Fe/H{]}\( \sim  \)\( - \)2.5. The data seem to indicate that {[}N/Fe{]}
remains more or less constant over the range of metallicities studied, but besides
the large dispersion, there are also uncertainties on the value of the constant,
although the data are compatible with a solar value. Our model predicts a rapid
increase of {[}N/Fe{]} at low metallicity, which show that metal poor massive
stars do not produce primary nitrogen. The ratio {[}N/Fe{]} steadily increases
up to {[}Fe/H{]}\( \sim -1 \) due to the progressive contribution of mostly
secondary nitrogen ejected by intermediate and low-mass stars, until the iron
production by Type Ia supernovae compensates the ejecta of intermediate stars,
flattening the evolution of {[}N/Fe{]}.

TWW\cite{timm95} obtained primary nitrogen production by metal poor massive
stars with masses above 30 M\( _{\sun } \) by considering enlarged convective
overshooting, but still within the theoretical uncertainties, due to the violent
effects of the penetration of the helium convective burning shell into the hydrogen
shell, although this effect does not occur in more metal-rich massive stars.
This could be a promising mechanism to explain the abundance pattern of damped
Lyman-\( \alpha  \) systems (Matteucci et al. \cite{matt97}). Synthesis of
primary nitrogen due to the injection of protons into helium burning zones could
also result in stellar models that include rotation (Heger et al. \cite{hege99};
Maeder \& Meynet \cite{maed00}). Massive stars produce \element[][15]{N} but
the dominant contribution in our model comes from novae.

Oxygen is exclusively produced by massive stars and dominates their ejecta.
Until recent years the observational status of the oxygen abundance in dwarfs
and G and K giants showed a nearly constant value of {[}O/Fe{]}\( \sim  \)0.5
for {[}Fe/H{]}\( \leq  \)\( - \)1, and a gradual decline in the disk. In fact,
this is the canonical behavior of the so-called \( \alpha  \)-elements (O,
Mg, Si, S Ca, Ti): almost flat evolution in the halo, and a gradual decline
in the disk due to the iron production by Type Ia supernovae. However, recent
data by Israelian et al. (\cite{isra98}) and Boesga1ard et al. (\cite{boes99}),
which agree with earlier results from Abaia \& Rebolo (\cite{abia89}), are
in contradiction with the idea of an oxygen plateau, since they find that {[}O/Fe{]}
increases between {[}Fe/H{]}=\( - \)1 and -3, from 0.6 to 1, which means a
slope in this range of -0.31\( \pm  \)0.11. 

Our calculations show that {[}O/Fe{]} begins with a value of the order of 1
at very low metallicity, and it slowly declines during the thick disk phase,
in part because massive stars of different masses and initial metal contents
produce different O/Fe ratios, and also because Type Ia supernovae begin to
inject iron already in the halo-thick disk phase, though their influence is
stronger during the thin disk epoch, when Type Ia supernovae reach their highest
rate, as it is clearly shown by the steeper decline of {[}O/Fe{]} for {[}Fe/H{]}\( \geq  \)-1.
This behaviour is in agreement with previous results by TWW\cite{timm95} and
Chiappini et al. (\cite{chia97}). We obtain a slope of -0.28, which clearly
indicates a deviation from the traditional oxygen plateau, although a little
bit less pronounced than in the new data. We will see the same kind of evolution,
although less marked, for the rest of the \( \alpha  \)-elements.

\subsection{Fluorine}

In our calculations fluorine is synthesized (together with \element[][7]{Li}
and \element[][11]{B} which will not be treated here) by the so-called neutrino-induced
nucleosynthesis in Type II supernova explosions. The amount of fluorine produced
by neutrino spallation of \element[][20]{Ne} is extremely sensitive to the neutrino
fluxes and spectra used in the explosion calculations, which render its yield
very uncertain.

The upper panel of Fig. \ref{abunF} displays the evolution versus iron abundance
of the ratio {[}F/Fe{]}. Our model predicts a practically constant and almost
solar {[}F/Fe{]}. In the lower panel of Fig. \ref{abunF} we present the ratio
{[}F/O{]} compared to the only available set of observations in the literature.
In agreement with TWW\cite{timm95}, our model predicts strong subsolar {[}F/O{]}
ratios at low metallicity, which is a result of the early efficient oxygen enrichment
by massive stars and not of a deficient production of fluorine, while it fits
the observed abundances of normal K giants.

Most of the plotted data correspond to spectroscopically peculiar giant stars
whose abundances are affected by stellar evolution, although the high ratios
observed in those stars also point to contributions other than from massive
stars. Forestini \& Charbonnel (\cite{fore97}) suggest fluorine production
in the helium burning shell of AGB stars, and Meynet \& Arnould (\cite{meyn00})
contributions from WR stars. A reliable history of fluorine will only be settled
when we gain new insight on the contributions from different kinds of stars.

\subsection{The \protect\( \alpha \protect \)-elements Magnesium, Silicon, Sulphur and
Calcium}

\begin{figure}
{\par\centering \resizebox*{\hsize}{!}{\includegraphics{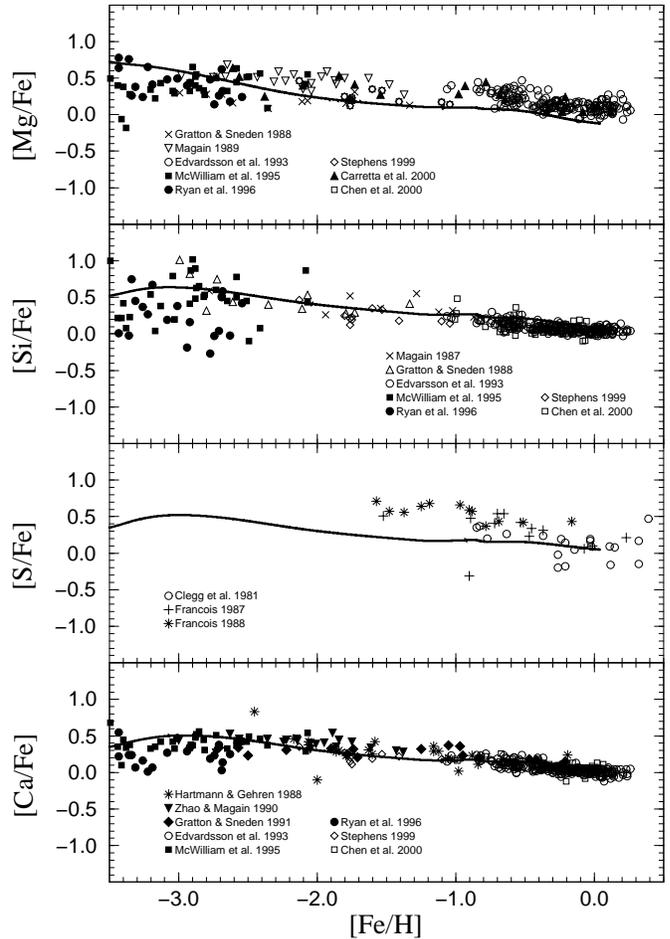}} \par}

\caption{Evolution of the abundance ratios to iron of the \protect\( \alpha \protect \)
elements Mg, Si, S, and Ca as a function of {[}Fe/H{]}.}
\label{abunMgSiSCa}
\end{figure}

Our results for the evolution of the ratios {[}\( \alpha  \)/Fe{]} for magnesium,
silicon, sulphur and calcium, plotted in Fig. \ref{abunMgSiSCa}, closely follow
the general trends of the data. Chiappini et al. (\cite{chia99}) pointed out
that the so-called ``plateau'' for the \( \alpha  \)-elements at low metallicities
is not perfectly flat, but instead presents a slight slope. In fact, that is
exactly what we find: the abundance ratios slowly increase with decreasing {[}Fe/H{]},
in good agreement with the work of McWilliam et al. (\cite{mcwi95}) and that
of Ryan et al. (\cite{ryan96}) for magnesium, silicon and calcium, although
the slopes are shallower than for oxygen.

The results for silicon and calcium are in excellent agreement with the data.
The evolution of magnesium and sulphur, however, is less satisfactory.

In the case of magnesium, though we begin with typical halo values, the rapid
decline of {[}Mg/Fe{]} gives values a little lower than observed in the thick
disk phase, but still marginally compatible with the observations. Later on,
when the thin disk starts, the iron from Type Ia supernovae worses the problem
and we clearly underestimate the {[}Mg/Fe{]} ratio in the disk. In principle
the metallicity-independent yields for massive stars from Thielemann et al.
(\cite{thie96}) or the WW\cite{woos95} yields for solar metallicity could
solve this discrepancy (Thomas et al. \cite{thom98}; Chiappini et al. \cite{chia99}),
but this type of yields are not appropriate when studying the galactic halo.
Since the magnesium yields of WW\cite{woos95} increase with the stellar mass,
the use of the rather steep IMF of Kroupa et al. (\cite{krou93}) worsen the
situation. As pointed out by GP\cite{gosw00}, the use of the metallicity dependent
yields of Limongi et al. (\cite{limo00}) will probably solve the underproduction
of magnesium but the ratios \( \alpha  \)/Mg will not reproduce the observations.
The fact that current yields from massive stars cannot completely explain the
magnesium evolution could indicate the need of a supplementary source for this
element, either in low and intermediate mass stars or by enhancing the magnesium
yield in Type Ia supernovae (TWW\cite{timm95}).

The paucity of data for sulphur at low {[}Fe/H{]} does not allow precise comparisons,
but if the points by François (\cite{fran87}, \cite{fran88}) around and below
{[}Fe/H{]}\( \sim  \)-1 adequately represent the main trend, a supplementary
production of sulphur at low metallicity seems necessary. Given our nucleosynthetic
sources, sulphur is only produced by massive stars, and the uncertainties in
the WW\cite{woos95} yields of metal poor massive stars could be the cause of
the low {[}S/Fe{]} at low {[}Fe/H{]}. It has been suggested, however, that the
{[}S/Fe{]} values measured in the halo stars should be reduced by 0.2 dex (Lambert
\cite{lamb89}). If that were the case, our results for sulphur would nicely
fit the corrected data.

\subsubsection{The Magnesium isotopes}

Unlike most of the chemical elements, in the case of magnesium the evolution
of the isotopic ratios can be confronted with observational results.

\begin{figure}
{\par\centering \resizebox*{\hsize}{!}{\includegraphics{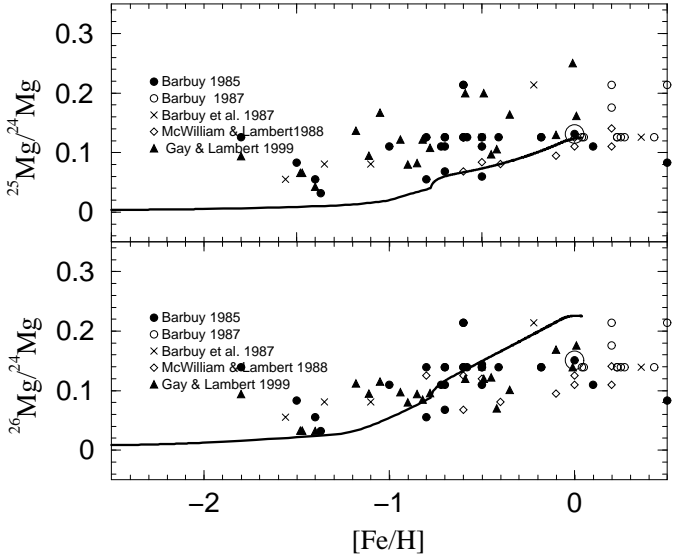}} \par}

\caption{Evolution of the magnesium isotopic ratios \element[][25]{Mg}/\element[][24]{Mg}
(\textit{upper panel}) and \element[][26]{Mg}/\element[][24]{Mg} (\textit{lower
panel}) as a function of metallicity.}
\label{Mgisot}
\end{figure}

In the upper panel of Fig. \ref{Mgisot} we show our results for the evolution
of the ratio \element[][25]{Mg}/\element[][24]{Mg} as a function of {[}Fe/H{]},
while in the lower panel appears the evolution of the ratio \element[][26]{Mg}/\element[][24]{Mg}.
The three isotopes are produced during the hydrostatic evolution of massive
stars. Since in the WW\cite{woos95} yields for massive stars \element[][24]{Mg}
is basically a primary isotope, while the neutron rich isotopes \element[][25]{Mg}
and \element[][26]{Mg} increase with metallicity (both are affected by the neutron
excess), it is to be expected, indeed, that both ratios decrease towards lower
metallicities. 

Our results, similar to those found by TWW\cite{timm95} and by GP\cite{gosw00},
follow the expected trend, but the model produces isotopic ratios clearly below
the observations in the halo-thick disk phase (i.e., {[}Fe/H{]}\( \leq  \)-1).
If the WW\cite{woos95} yields do not underestimate the magnesium isotopic yields,
a supplementary source of \element[][25]{Mg} and \element[][26]{Mg} is needed,
for instance through s-process in AGB stars (Iben \& Renzini \cite{ibe83}).

In the thin disk phase ({[}Fe/H{]} \textgreater{} -1), the model reproduce well
the global trend of the observations, with both ratios systematically increasing
with metallicity. The ratio \element[][25]{Mg}/\element[][24]{Mg} reaches the
solar value. However, the \element[][26]{Mg}/\element[][24]{Mg} ratio at {[}Fe/H{]}=0
is \( \sim  \)50\% higher than solar, again in agreement with the results of
GP\cite{gosw00} who attribute the overabundance of \element[][26]{Mg} at the
solar birth to the effects of the IMF of Kroupa et al. (\cite{krou93}) that
favor the production of \element[][26]{Mg} in front of that of \element[][24]{Mg}.

\subsection{Sodium, Aluminum and Potassium}

Sodium, aluminum and potassium are odd elements (besides, sodium and aluminum
are monoisotopic) mainly produced by massive stars. They are subjected to the
odd-even effect, i.e., stellar yields increasing with metallicity. Therefore,
their galactic abundance ratios to iron must increase with time, although the
evolution of the ratio {[}element/Fe{]} could be affected by the iron production
by Type Ia supernovae, especially in the disk. One can eliminate the effects
of Type Ia supernovae on the abundance evolution of odd elements by plotting
the ratio {[}element/Mg{]}, as in TWW\cite{timm95}, because magnesium is also
nearly exclusively synthesized in massive stars, or by using as metallicity
indicator an element more reliable than iron, which is affected by uncertainties
such as the evolution of Type Ia supernova rates or the mass cut and explosion
energy in Type II supernovae, as done by GP\cite{gosw00} who use calcium instead
of iron as metallicity indicator.

\begin{figure}
{\par\centering \resizebox*{\hsize}{!}{\includegraphics{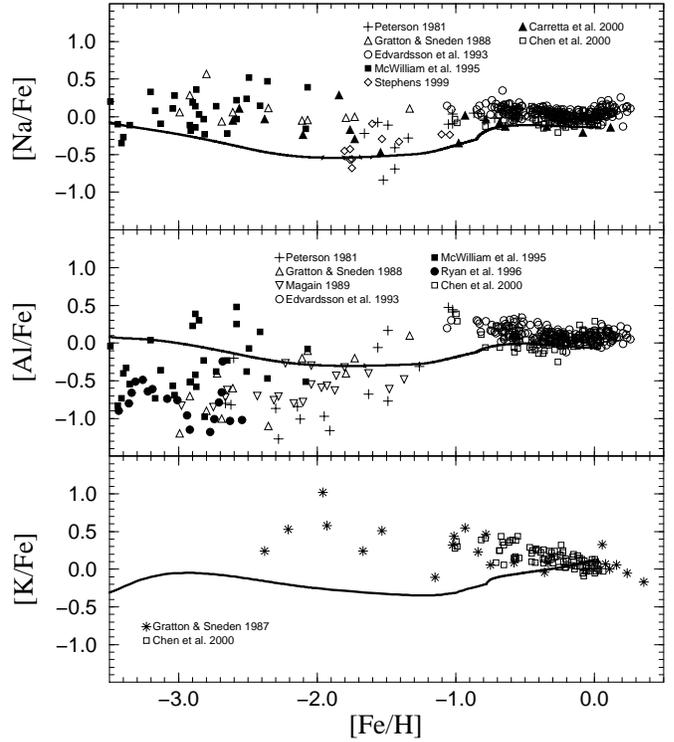}} \par}

\caption{Evolution of Na, Al and K abundance ratios to iron in terms
of {[}Fe/H{]} as metallicity indicator. }
\label{abunNaAlK}
\end{figure}

\begin{figure}
{\par\centering \resizebox*{\hsize}{!}{\includegraphics{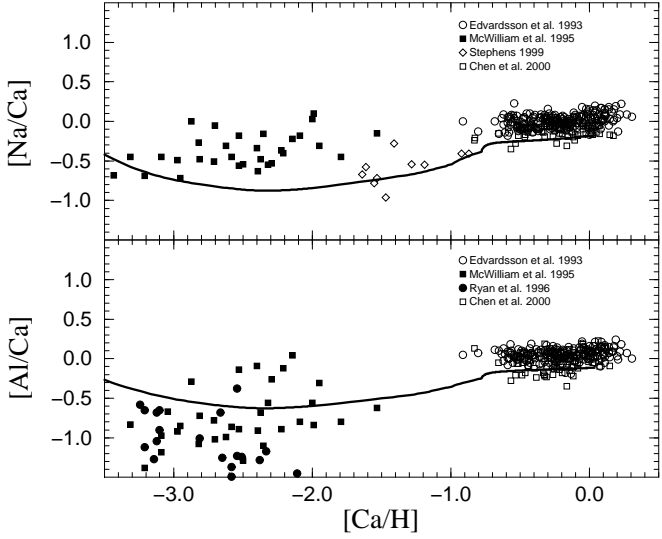}} \par}

\caption{Evolution of the ratios {[}Na/Ca{]} and {[}Al/Ca{]} in
terms of {[}Ca/H{]} as metallicity indicator.}
\label{abunNaAl/Ca}
\end{figure}

We display in Fig. \ref{abunNaAlK} the calculated {[}Na/Fe{]} (upper panel)
and {[}Al/Fe{]} (central panel). In Fig. \ref{abunNaAl/Ca} we also show the
ratios {[}Na/Ca{]} and {[}Al/Ca{]} using {[}Ca/H{]} as a measure of the metallicity.
It is worth to stress that the evolution of the calcium abundance obtained in
our model agrees very well with the data (see Fig. \ref{abunMgSiSCa}).

The observed sodium to iron ratios display a plateau in the disk. The situation
for the halo stars is less clear. The rather old data by Peterson (\cite{pete81})
indicate a rapid decrease from {[}Fe/H{]}\( =-1 \) to {[}Fe/H{]}\( =-2 \).
Recent observations by Stephens (\cite{step99}) show the same trend, but most
other observations point to a constant {[}Na/Fe{]}\( \sim  \)0, although with
large dispersion. We obtain an almost constant, but lower than the data, abundance
up to {[}Fe/H{]}\( \sim  \)-2.5, and a steep increase at the beginning of the
thin disk formation that ends in a plateau lower than observed.

The calculated evolution of {[}Al/Fe{]} begins almost flat and is clearly higher
than the data from halo stars. Later, it slowly increases at the thick-thin
disk transition, but, as with Na, remains below the disk star observations.

The abundance ratio to calcium of both elements shows now the odd-even effect
(see Fig.\ref{abunNaAl/Ca}). We still have low {[}Na/Ca{]} in the complete
range of {[}Ca/H{]}, but the aluminum ratio shows a better agreement with the
observations. We remind that our models underproduce the solar sodium by a factor
of \( \sim  \)1.4, and aluminum by a factor of \( \sim  \)1.2.

As pointed out by TWW\cite{timm95} and GP\cite{gosw00}, intermediate-mass
stars could produce some sodium and aluminum through the Ne-Na and Mg-Al cycles,
improving the fit to the data.

The potassium abundance is dominated by the odd isotope \element[][39]{K}, and
it is again mainly produced in massive stars through hydrostatic oxygen burning,
so that it should show the odd-even effect. Surprisingly, the observed abundances
display the opposite evolution: the few halo observations give values of {[}K/Fe{]}\( \sim  \)0.5
with a wide scatter, decreasing to solar values in the disk. Our results, presented
in Fig.\ref{abunNaAlK}, show, however, a behavior similar to other odd elements.
We obtain slightly lower than solar values in the halo-thick disk, and a gradual
rise during the thin disk phase towards final solar values, in agreement with
the reduced iron yield evolution of TWW\cite{timm95}.

\begin{figure}
{\par\centering \resizebox*{\hsize}{!}{\includegraphics{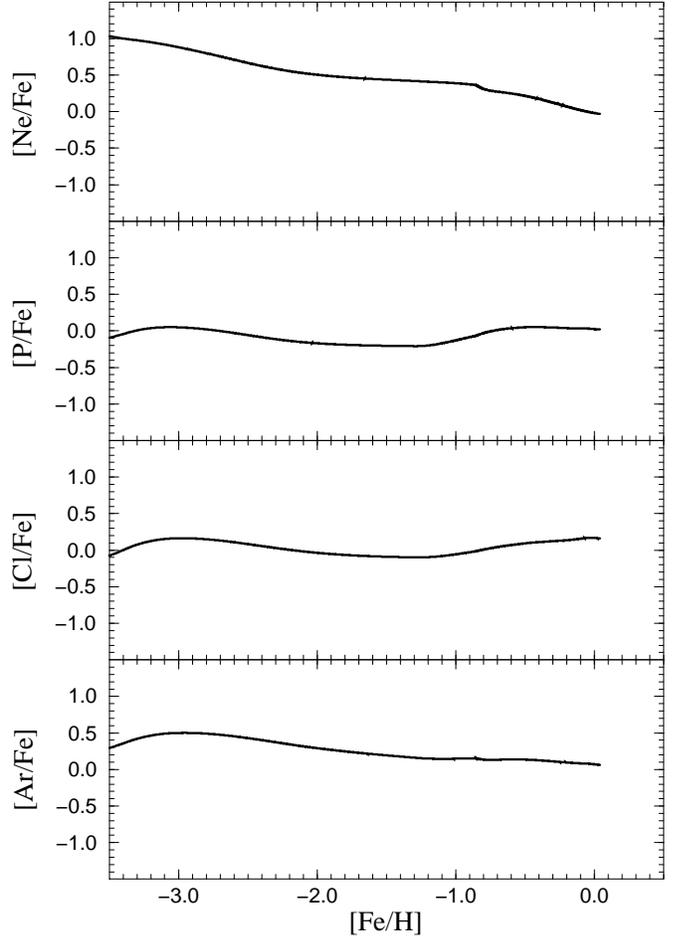}} \par}

\caption{Evolution of neon, phosphorus, chlorine and argon abundance
ratios to iron, as a function of {[}Fe/H{]}.}
\label{abunNePClAr}
\end{figure}

\subsection{Neon, Phosphorus, Chlorine and Argon}

Although there are not observational data to compare with, we plot in Fig. \ref{abunNePClAr}
the calculated evolution of the noble gases neon and argon, as well as those
of phosphorus and chlorine, just for completeness.

Neon abundance is dominated by \element[][20]{Ne}, and the three argon isotopes
are also of even Z. Therefore, both elements should then behave as \( \alpha  \)-elements.
The calculations confirm this theoretical expectation. The evolution of neon
in the halo-thick disk shows a clear dependence on metallicity, reminding of
the oxygen behavior. For argon the metallicity effect is less marked, and its
evolution remind that of intermediate \( \alpha  \)-elements like silicon or
calcium. In the absence of stellar observations for these elements, it is gratifying
that their solar abundances are well reproduced.

Both odd-Z elements, phosphorus and chlorine, present solar or slightly subsolar
abundance ratios in the halo-thick disk. When the assembling of the thin disk
begins, their abundances gradually climb towards the solar values due to the
odd-even effect in massive stars (and in a much lesser extent, by contribution
of Type Ia supernovae), until the iron from Type Ia supernovae flattens the
evolution of phosphorus, although it is not the case for chlorine.

\begin{figure}
{\par\centering \resizebox*{\hsize}{!}{\includegraphics{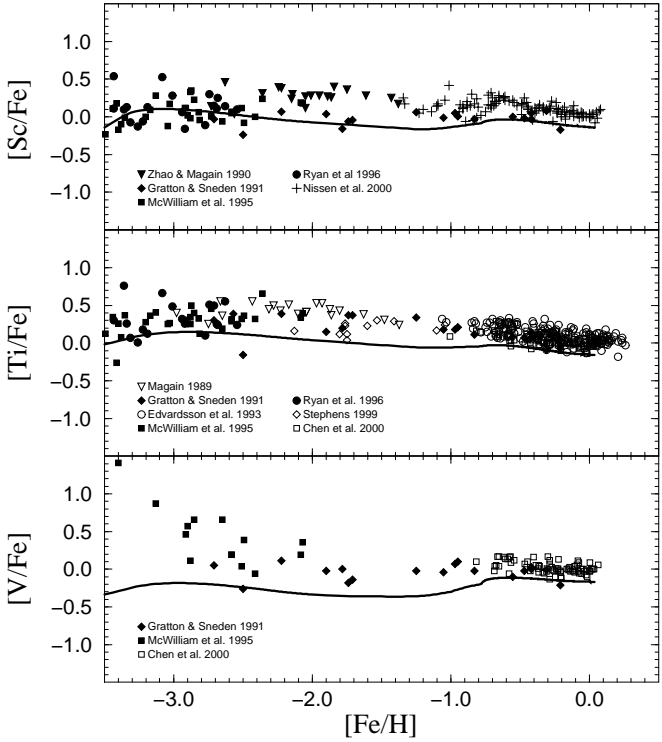}} \par}

\caption{Evolution with metallicity of the iron peak elements scandium,
titanium and vanadium as a function of {[}Fe/H{]}.}
\label{abunScTiV}
\end{figure}

\subsection{The Iron Peak elements Scandium, Titanium, Vanadium, Chromium, Manganese, Cobalt
and Nickel}

All these elements are basically synthesized through explosive burning in Type
II and/or Type Ia supernova explosions, and some of them can also be produced
by neutron captures during hydrostatic He and C burning. Their stellar yields
are therefore much more uncertain than those of the intermediate elements studied
above due to our deficient understanding of the parameters that determine supernova
explosions (energy and mass cut in Type II supernovae, progenitors, burning
propagation and rates of Type Ia supernovae), and of neutron fluxes in hydrostatic
He and C combustion. Isotopes with A \textless{} 56 are mainly produced in explosive
O and Si burning, and in Nuclear Statistical Equilibrium (NSE), while isotopes
with A\( \geq  \)56 are produced in alpha-rich freeze-out NSE, besides contributions
from neutron captures.

\begin{figure}
{\par\centering \resizebox*{\hsize}{!}{\includegraphics{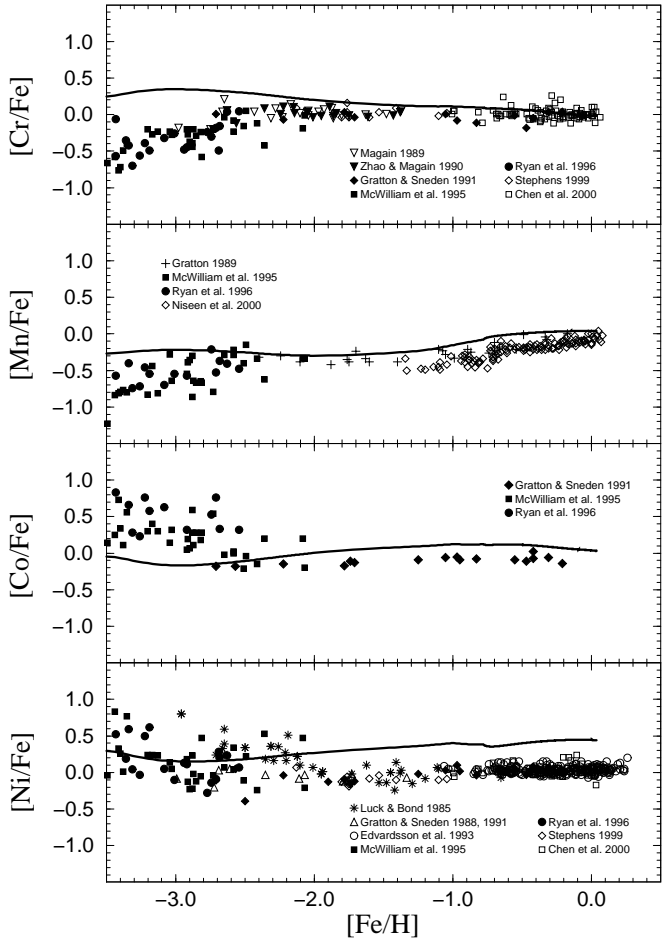}} \par}

\caption{Evolution of the iron peak elements Cr, Mn, Co and Ni
as a function of {[}Fe/H{]}}
\label{abunCrMnCoNi}
\end{figure}

The calculated history of scandium, titanium and vanadium is displayed in Fig.
\ref{abunScTiV}, and that of chromium, manganese, cobalt and nickel appear
in Fig. \ref{abunCrMnCoNi}. 

The evolution for the {[}Sc/Fe{]} ratio agrees reasonably well with the data
for halo stars, characterized by a flat and nearly solar value, although between
{[}Fe/H{]} = -2.5 and {[}Fe/H{]} = -1 we get a gentle decrease and our results
fall below the observations. At the transition to the thin disk the abundance
grows a bit, as expected for an odd element, but iron from Type Ia supernovae,
which make little scandium, overwhelms this increase and lowers the ratio {[}Sc/Fe{]}
during the disk evolution, which is still below solar.

Titanium, dominated by \element[][48]{Ti}, should behave as an \( \alpha  \)-element.
In fact, the observations point to {[}Ti/Fe{]}\( \sim  \)0.3-0.4 in the most
metal-poor halo stars, and to a clear trend towards a declining ratio as metallicity
increases. Loosely speaking, our calculations reproduce such characteristics,
but the abundance is systematically lower than observed for the complete range
of metallicity. At the solar birth all the titanium isotopes except \element[][46]{Ti}
are underabundant (see Fig. \ref{XXsun}). TWW\cite{timm95} and GP\cite{gosw00}
found the same problems with titanium. The conclusion that current yields from
massive stars alone fail to explain the titanium history is unavoidable.

The isotope \element[][51]{V} is predominant in the total abundance of vanadium.
The few available data show a {[}V/Fe{]} flat and solar in population I and
in metal-rich halo stars, and a tendency to increase as {[}Fe/H{]} decreases
from -2 to -3. The calculations give a flat {[}V/Fe{]} in the halo-thick disk
with a value of \( \sim  \)-0.3, a moderate and rapid rise at the beginning
of the thin disk formation, and a flat evolution slightly lower than solar during
most of the thin disk phase; here, again, iron from Type Ia supernovae compensates
the vanadium produced by massive stars with Z \( \geq  \) 0.1Z\( _{\sun } \).

The observations on the ratio {[}Cr/Fe{]} available until mid-nineties showed
a constant and solar value down to {[}Fe/H{]}\( \sim  \)-2.5, i.e. the same
behavior as iron, indicating that both elements were produced in the same scenarios.
However, recent observations show that the chromium abundance ratio to iron
decreases as the metallicity goes down from {[}Fe/H{]} =-2.5 to {[}Fe/H{]} =
-3.5, reaching at the latter metallicity values as low as {[}Cr/Fe{]}\( \sim  \)-0.7.
The chromium stellar yields of WW\cite{woos95} are quite insensitive to Z.
Our results, shown in Fig. \ref{abunCrMnCoNi}, give an almost flat {[}Cr/Fe{]}
in the complete range of metallicities, and thus fail to reproduce the data
at low metallicity. The flatness of the abundance ratio is more marked in the
disk because chromium and iron are produced in the same proportions by Type
Ia supernovae, fitting precisely the solar abundances of the chromium isotopes. 

In the case of manganese, with only one stable and odd-Z isotope (\element[][55]{Mn}),
the calculated evolution fits well the data for {[}Fe/H{]}\textgreater{}-2.5,
although in the thin disk there is an slight overabundance. The roughly linear
increase in the disk reveals the odd-even effect for this element. At very low
metallicities, as for chromium, the calculated abundances lie clearly above
the data.

The observational situation for the ratio {[}Co/Fe{]} is rather puzzling. Cobalt
is an odd element and therefore one expects its abundance to diminish with decreasing
metallicity. Instead, the latest data point to a noticeable increase of its
abundance at the lowest metallicities. We got a rough agreement down to {[}Fe/H{]}
\( =- \)2.5, where {[}Co/Fe{]} mildly decline when going to lower metallicities,
although the calculated history appears to lie higher than the observations.
Below {[}Fe/H{]}\( =- \)2.5, the abundance ratio continues its gentle decline,
just the opposite trend to that observed.

Nickel abundances in halo and disk stars have been obtained in a large number
of surveys. Most of them show a constant and solar value for the ratio {[}Ni/Fe{]}
at all metallicities, but the most recent observations seem to indicate an increasing
ratio towards low {[}Fe/H{]}, although the scatter is really large. The models
give a continuous increase of an overabundant {[}Ni/Fe{]} for {[}Fe/H{]}\textgreater{}-3.
We obtain higher values than observed due to the combination of a reduction
of a factor of two in the WW\cite{woos95} iron yields from massive stars, together
with the large amount of \element[][58]{Ni} in the yields of the W7 model of
Type Ia supernovae, the latter being responsible for the positive slope in the
thin disk.

\subsection{Beyond the iron peak: Copper and Zinc}

\begin{figure}
{\par\centering \resizebox*{\hsize}{!}{\includegraphics{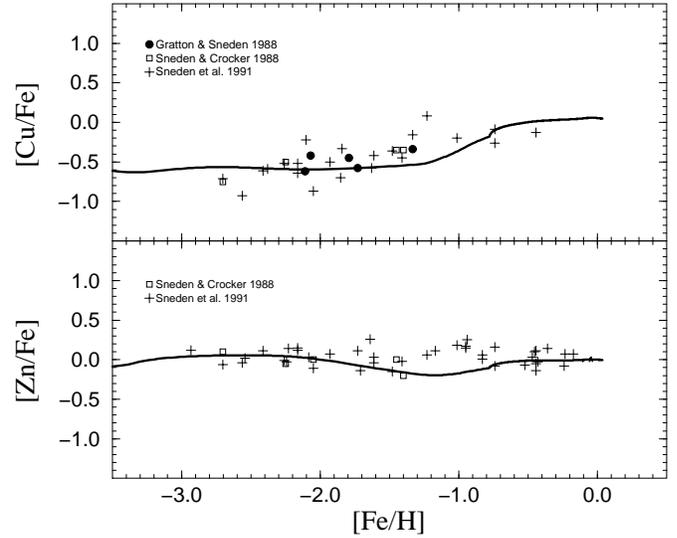}} \par}

\caption{Evolution of copper and zinc as a function of {[}Fe/H{]}}
\label{abunCuZn}
\end{figure}

These two elements are produced in significant amounts at the different burning
stages in massive stars. As for the iron peak elements, the total ejected amounts
of these nuclei are strongly dependent on the characteristics of the supernova
explosions. Nevertheless, ratios relative to iron could be less affected by
astrophysical uncertainties. For instance, as shown by WW\cite{woos95}, in
the explosion of a solar metallicity 35 M\( _{\sun } \) model, variations by
a factor 2 in the explosion energy give changes in the {[}Ni/Fe{]} ratio of
less than 10\%. This argument in favor of the relatively robustness of the calculated
evolution of the element ratios relative to iron also applies to copper and
zinc. The evolution of both elements is presented in Fig. \ref{abunCuZn}. 

Observations for these elements are scarce because they do not present enough
suitable transitions. According to the available data, copper behaves like a
typical odd nucleus, with a marked odd-even effect which gives a continuous
growth of {[}Cu/Fe{]} with the iron contents. Our model nicely fits the data.
At low metallicities we obtain a flat evolution, and in the thin disk the odd-even
effect is clearly seen. As shown in Fig. \ref{XXsun}, the solar abundances
of the two copper isotopes are well adjusted.

The data for the zinc abundance show that this element evolves as iron, with
a constant and solar {[}Zn/Fe{]} in the complete interval of observed metallicities.
Here, again, the calculated evolution fits well the data. However, this result
could be misleading since even if we obtain an elemental abundance in almost
perfect agreement with the solar value, the distribution between the four isotopes
of zinc (see Fig. \ref{XXsun}) does not agree with the solar isotopic ratios.
As mentioned above, we underproduce by \( \sim  \) a factor of 2 the most abundant
isotope, \element[][64]{Zn}, and overproduce \element[][68]{Zn} by the same
factor.

\section{Summary and conclusions}

We have analyzed the evolution of all the stable isotopes between hydrogen and
zinc (lithium, beryllium and boron isotopes will be treated in a future paper)
by means of a detailed chemical evolution code. In the framework of standard
chemical evolution models only the works by TWW\cite{timm95} and by GP\cite{gosw00}
have studied such a large number of isotopes. In the first paper, the model
employed was adequate just for the galactic disk evolution, apart from using
the rather old yields from Renzini \& Voli (\cite{renz81}) for low and intermediate-mass
stars. In the second case, even if the halo and the disk were correctly treated,
only yields from massive stars were included

Our model assumes the formation of the Milky Way in two main episodes of exponentially
decreasing infall. The first episode forms the halo and the thick disk from
infall of primordial extragalactic material with an e-folding timescale of 1
Gyr. Then, the thin disk is assembled by infall of metal rich extragalactic
gas with a global metallicity of 0.1Z\( _{\sun } \) in solar proportions, and
with a relatively long timescale (7 Gyr at the solar ring). We have also calculated
models that accrete material of primordial composition during the complete evolution.
Although the results of the two kind of models are similar, we obtain a better
agreement with the data for the stellar distribution metallicities in the case
of the model with enriched infall. We have employed an up to date IMF (the version
of Kroupa et al. (\cite{krou93})), and a form for the SFR (the Dopita \& Ryder's
(\cite{dopi94}) law), that relies on observations of the properties of star
formation in spiral galaxies.

One important characteristic of our model is that it incorporates metallicity
dependent yields for isolated stars in the whole range of stellar masses. We
use the WW\cite{woos95} yields for massive stars and van der Hoek \& Groenengen
(\cite{vand97}) for low and intermediate mass stars. Following TWW\cite{timm95},
in order to obtain a better concordance with observations, we have reduced the
rather uncertain iron yield of WW\cite{woos95} by a factor of two. We have
not included metallicity effects in the yields of Type Ia supernovae, but GP\cite{gosw00}
pointed out that the inclusion of the metallicity dependent yields of Iwamoto
et al. (\cite{iwam99}) hardly affects the final results. Finally we have taken
into account the nucleosynthetic contributions from novae that give important
contributions to isotopes such as \element[][13]{C}, \element[][15]{N}, and
\element[][17]{O}.

We have compared in detail the theoretical results with a large ensemble of
observational data. First of all, in Sec. 3 we have shown that our results are
able to reproduce the main characteristics of the solar vicinity, especially
the G-dwarf distribution of the local disk, that we consider the most stringent
restriction. We also obtain a really nice agreement with the solar composition
(both elemental and isotopic) at the time of 8.5 Gyr. It is worth to stress
that:

\begin{itemize}
\item All the solar elemental abundances are reproduced within a factor of two, except
for nickel, which is slightly oversolar due to the large production factor of
\element[][58]{Ni}i in the adopted yields of Type Ia supernovae and of \element[][62]{Ni}
in the yields of exploded massive stars of WW\cite{woos95}. 
\item The contributions from low and intermediate-mass stars suffice to build the
abundances of the main CNO isotopes. The inclusion of novae increases the solar
abundances of \element[][13]{C} (by a factor of \( \sim  \)1.6), and noticeably
that of \element[][15]{N} (by a factor of \( \sim  \)4) and \element[][17]{O}
(by a factor of \( \sim  \)2).
\item Massive star yields still encounter difficulties to account for the underproduction
of the neutron-rich isotopes \element[][48]{Ca} and \element[][50]{Ti}, as it
is also the case with \element[][47]{Ti}.
\end{itemize}
The evolution of the abundance ratios {[}element/Fe{]} obtained in our model
has been confronted with a large and updated set of observations. The main conclusions
that result from the comparison are:

\begin{itemize}
\item The evolution of the CNO elements is very well reproduced by our model. As it
is well known, massive stars underproduce carbon and nitrogen (TWW\cite{timm95};
GP\cite{gosw00}). However, the inclusion of low and intermediate mass stars
is enough to reproduce adequately the carbon evolution and its solar abundance.
The behavior of nitrogen is again well reproduced, although some primary nitrogen
from massive stars could be necessary during the halo phase. For oxygen we obtain
a clear deviation of the so-called \( \alpha  \)-plateau which is also present,
in a lesser extent, in the rest of the \( \alpha  \)-elements . Our results
show a continuous decrease of the ratio {[}O/Fe{]} from the beginning of the
halo-thick disk phase, as pointed out first by TWW\cite{timm95} and then by
Chiappini et al. (\cite{chia99}). The slope is still shallower than in the
data from Israelian et al. (\cite{isra98}) and Boesgaard et al. (\cite{boes99}),
but the fit with the more recent data of Carreta et al. (\cite{carr00}) is
excellent.
\item As with oxygen, most of the \( \alpha  \) and \( \alpha  \)-like elements
show some deviations of the typical \( \alpha  \)-plateau, in the sense that
their abundances ratios to iron continuously decrease with increasing metallicity
due to the combined effects of stellar yields depending on mass and metallicity
and the contribution to the iron contents for Type Ia supernovae, in agreement
with observations. However, some of these elements are underproduced at all
metallicities by the WW\cite{woos95} yields (as well as by the yields from
Limongi et al. (\cite{limo00}), according to GP\cite{gosw00}), especially
magnesium and titanium.
\item In general, the intermediate odd-Z elements show the expected theoretical behaviour
as a consequence of the marked odd-even effect in the stellar yields. However,
for several of these elements there exist important discrepancies with the observational
data, in particular for sodium and potassium. The same discrepancy appears in
the case of scandium and vanadium which, besides, are underproduced by the model
at all metallicities.
\item Although to zero order the evolution of the iron peak elements is consistent
with the data, there are noticeable discrepancies. Chromium and manganese are
overabundant at low metallicities, which may indicate problems with the iron
yields for stars of low metallicity, while nickel is overabundant at moderate
and high metallicity. This last flaw has to be attributed to the nickel yields
of model W7 for Type Ia supernovae. In any case, as mentioned above, the calculations
of massive star explosions are still plagued with severe problems.
\end{itemize}


\begin{thebibliography}{1989}
\bibitem[1989]{abia89}Abia, C., Rebolo, R., 1989, ApJ 347, 186
\bibitem[1989]{ande89}Anders E., Grevesse N., 1989, Geochim. Cosmochim. Acta 53, 197
\bibitem[1985]{barb85}Barbuy B., 1985, A\&A 151, 189
\bibitem[1987]{barb87a}Barbuy B., 1987, A\&A 172, 251
\bibitem[1989]{barb89}Barbuy B., Erdelyi-Mendes M., 1989, A\&A 214, 239
\bibitem[1987]{barb87b}Barbuy B., Spite F., Spite M., 1987, A\&A 178, 199
\bibitem[1999]{berc99}Berczik P., 1999, A\&A 348, 371
\bibitem[1999]{boes99}Boesgaard A.M., King J.R., Deliyannis C.P., Vogt S.S., 1999, AJ 117, 492
\bibitem[1999]{bois99}Boissier S., Prantzos N., 1999, MNRAS 307, 857
\bibitem[1992]{burk92}Burkert A., Truran J.W., Hensler G., 1992, ApJ 391, 651
\bibitem[1998]{buse98}Buser R., Rong J., Karaali S., 1998, A\&A 331, 934
\bibitem[1987]{carb87}Carbon D.F., Barbuy B., Kraft R.P., Friel E.D., Suntzeff N.B., 1987, PASP 99,
335
\bibitem[1997]{card97}Cardelli J., Federman S., 1997, in Nuclei in the Cosmos IV, ed. J. Gorres et
al. (Amsterdam: Elsevier), 31
\bibitem[1994]{cari94}Carigi L., 1994, ApJ 424,181
\bibitem[2000]{carr00}Carretta E., Gratton R.G., Sneden C., 2000, A\&A 356, 238
\bibitem[1999]{chan99}Chang R.X., Hou J.L., Shu C.G., Fu C.Q., 1999, A\&A 350, 38
\bibitem[1996]{char96}Charbonnel C., Meynet G., Maeder A., Schaerer D., 1996, A\&AS 115, 339
\bibitem[2000]{chen00}Chen Y.Q., Nissen P.E., Zhao G., Zhang H.W., Benoni T., 2000, A\&AS 141, 491
\bibitem[1997]{chia97}Chiappini C., Matteucci F., Gratton G., 1997, ApJ 477, 765
\bibitem[1999]{chia99}Chiappini C., Matteucci F., Beers T.C., Nomoto K., 1999, ApJ 515, 226
\bibitem[2000]{chia00}Chiappini C., Matteucci F., Padoan P., 2000, ApJ 528, 711
\bibitem[1981]{cleg81}Clegg R.E.S., Tomkin J., Lambert D. L., 1981, ApJ 250, 262
\bibitem[1993]{dick93}Dickey J.M., 1993, in ASP Conf. Ser. 39, The Minnesota Lectures on the Structure
and Dynamics of the Milky Way, ed. R. M. Humphreys (San Francisco: ASP), 93
\bibitem[1994]{dopi94}Dopita M.A., Ryder S.D., 1994, ApJ 430, 163
\bibitem[1984]{edmu84}Edmunds M.G., Pagel B.E.J., 1984, MNRAS 211, 507
\bibitem[1993]{edva93}Edvardsson B., Anderson J., Gustafsson B., Lambert D.L., Nissen P.E., Tomkin
J., 1993, A\&A 275, 101
\bibitem[1999]{flyn99}Flynn C., Gould A., Bahcall J.N., 1999, in ASP Conf. Ser. 165, The Third Stromlo
Symposium: The Galactic Halo, ed. B.K. Gibson, T.S. Axelrod, M.E. Putman (San
Francisco: ASP), 387
\bibitem[1997]{fore97}Forestini M., Charbonnel C., 1997, A\&AS 123, 241
\bibitem[1987]{fran87}François P., 1987, A\&A 176, 294
\bibitem[1988]{fran88}François P., 1988, A\&A 195, 226
\bibitem[2000]{garn00}Garnett D.R., Kobulnicky H.A., 2000, ApJ 532, 1192
\bibitem[2000]{gay 00}Gay P.L., Lambert D.L., 2000, ApJ 533, 260
\bibitem[1989]{gilm89}Gilmore G., Wyse R., Kuijen K., 1989, in Evolutionary Phenomena in Galaxies,
J. Beckman, \& B. Pagel (eds.), Cambridge University Press, 172
\bibitem[1995]{giov95}Giovanoli A., Tosi M., 1995, MNRAS 273, 499
\bibitem[2000]{gosw00}Goswami A., Prantzos N., 2000, A\&A 359, 191 (GP2000)
\bibitem[1985]{grat85}Gratton R.G., 1985, A\&A 148, 105
\bibitem[1989]{grat89}Gratton R.G., 1989, A\&A 208, 171
\bibitem[1986]{grat86}Gratton R.G., Ortolani S., 1986, A\&A 169, 201
\bibitem[1987]{grat87}Gratton R.G., Sneden C., 1987, A\&A 178, 179
\bibitem[1988]{grat88}Gratton R.G., Sneden C., 1988, A\&A 204, 193
\bibitem[1991]{grat91}Gratton R.G., Sneden C., 1991, A\&A 241, 501
\bibitem[1982]{gust82}Güsten R., Mezger M., 1982, Vistas in Astr. 26, 159
\bibitem[1988]{hart88}Hartmann K., Gehren T., 1988, A\&A 199, 269
\bibitem[1997]{hata97}Hatano K., Branch D., Fisher A., Starrfield S., 1997, MNRAS 290, 113
\bibitem[1999]{hege99}Heger A., Langer N., Woosley S.E., 1999, ApJ 528, 368
\bibitem[1999]{henr99}Henry R.B.C., Worthey G., 1999, PASP 111, 919 
\bibitem[1983]{ibe83}Iben I., Renzini A., 1983, ARA\&A 21, 271
\bibitem[1988]{isra98}Israelian G., García López R., Rebolo R., 1998, ApJ 507, 805
\bibitem[1999]{iwam99}Iwamoto K., Brachwitz F., Nomoto K., Kishimoto N., Hix R., Thielemann K.-F,
1999, ApJS 125, 439
\bibitem[2000]{jorg00}J\o rgensen B. R., 2000, A\&A 363, 947
\bibitem[1992]{jori92}Jorissen A., Smith V.V., Lambert D.L., 1992, A\&A 261, 164
\bibitem[1998]{jose98}José J., Hernanz M., 1998, ApJ 494, 680
\bibitem[1998]{kenn98}Kennicutt R., 1998, ApJ 498, 541
\bibitem[1998]{krou98}Kroupa P., 1998, in Brown Dwarfs and Extrasolar Planets, R. Rebolo et al. (eds.),
ASP Conf. 134, 483
\bibitem[1993]{krou93}Kroupa P., Tout C., Gilmore G., 1993, MNRAS 262, 545
\bibitem[1989]{kuij89}Kuijken K., Gilmore G., 1989, MNRAS 239, 605
\bibitem[1985]{lair85}Laird J.B., 1985, ApJ 289, 556
\bibitem[1989]{lamb89}Lambert D.L., 1989, in Cosmic Abundances of Matter, ed. C.J. Waddington, AIP
Conf. 183, 168
\bibitem[1972]{lars72}Larson R.B., 1972, Nature 236, 21
\bibitem[2000]{limo00}Limongi M., Straniero O., Chieffi A., 2000, ApJS 129, 625
\bibitem[1985]{luck85}Luck R.E., Bond H.E., 1985, ApJS 59, 249
\bibitem[2000]{maed00}Maeder A., Meynet G., 2000, ARAA in press 
\bibitem[1987]{maga87}Magain P., 1987, A\&A 179, 176
\bibitem[1989]{maga89}Magain P., 1989, A\&A 209, 211
\bibitem[1996]{mari96}Marigo P., Bressan A., Chiosi C., 1996, A\&A 313, 545
\bibitem[1998]{mass98}Massey P., 1998, in The Stellar Initial Mass Function, ed. G.Gilmore, D. Howell.
ASP Conf. Ser. 142, 17
\bibitem[1995]{mass95}Massey P., Johnson K., Degioia-Eastwood K., 1995, ApJ 454, 151
\bibitem[1986]{matt86}Matteucci F., Greggio L., 1986, A\&A 154, 279
\bibitem[1997]{matt97}Matteucci F., Molaro P., Vladilo G., 1997, A\&A 321, 45
\bibitem[1988]{mcwi88}McWilliam A., Lambert D.L., 1988, MNRAS 230, 573
\bibitem[1995]{mcwi95}McWilliam A., Preston G.W., Sneden C., Searle L., 1995, AJ 109, 2757
\bibitem[1998]{mera98}Méra D., Chabrier G., Schaeffer R., 1998, A\&A 330, 937
\bibitem[1991]{meus91}Meusinger H., Reimann H.G., Stecklum B., 1991, A\&A 245, 57
\bibitem[2000]{meye00}Meyer M.R., Adams F.C., Hillenbrand L.A., Carpenter J.M., Larson R.B., 2000,
in Protostars and Planets IV (Book - Tucson: University of Arizona Press; eds
Mannings, V., Boss, A.P., Russell, S. S.), p. 121
\bibitem[2000]{meyn00}Meynet G., Arnould M., 2000, A\&A 355, 176
\bibitem[1981]{mira81}Mirabel I.F., 1981, Rev. Mex. Astron. \& Astrofis. 6, 245
\bibitem[1984]{mira84}Mirabel I.F., Morras R., 1984, ApJ 279, 86
\bibitem[2000]{mish00}Mishenina T.V., Korotin S.A., Klochkova V.G., Panchuk V.E., 2000, A\&A 353,
978
\bibitem[1994]{niss94}Nissen P.E., Gustafsson B., Edvardsson B., Gilmore G., 1994, A\&A 285, 440
\bibitem[2000]{niss00}Nissen P.E., Chen Y.Q., Schuster W.J., Zhao G., 2000, A\&A 353, 722
\bibitem[1984]{nomo84}Nomoto K., Thielemann F.-K., Yokoi K., 1984, ApJ 286, 644
\bibitem[1997]{page97}Pagel B., 1997, Nucleosynthesis and Galactic Chemical Evolution, Cambridge University
Press
\bibitem[1981]{pete81}Peterson R.C., 1981, ApJ 244, 989
\bibitem[1990]{pete90}Peterson R.C., Kurucz R.L., Carney B.W., 1990, ApJ 350, 173
\bibitem[1999]{port99}Portinari L., Chiosi C., 1999, A\&A 350, 827
\bibitem[1998]{port98}Portinari L., Chiosi C., Bressan A., 1998, A\&A 334, 505
\bibitem[1995]{pran95}Prantzos N., Aubert O., 1995, A\&A 302, 69
\bibitem[1998]{pran98}Prantzos N., Silk J., 1998, ApJ 507, 229
\bibitem[1992]{press92a}Press W.H., Teukolsky S.A., 1992, Comput. Phys. 6, 188
\bibitem[1992]{press92b}Press W.H., Teukolsky S.A., Vetterling W.T., Flannery B.P., 1992, Numerical
recipies in FORTRAN. The art of scientific computing, Cambridge University Press
\bibitem[1991]{rana91}Rana N.C., 1991, ARA\&A 29, 129
\bibitem[1997]{reid97}Reid I.N., Gizis J., 1997, AJ 113, 2246
\bibitem[1993]{reid93}Reid I.N., Majewski S.R., 1993, ApJ 409, 635
\bibitem[1981]{renz81}Renzini A., Voli M., 1981, A\&A 94, 175
\bibitem[1996]{robi96}Robin A.C., Haywood M., Creze M., Ojha D.K., Bienayme O., 1996, A\&A 305, 125
\bibitem[1996]{roch96}Rocha-Pinto H.J., Maciel W.J., 1996, MNRAS 279, 447
\bibitem[2000]{roch00}Rocha-Pinto H.J., Maciel W.J., Scalo J., Flynn C., 2000, A\&A 358, 850
\bibitem[2000]{roma00}Romano D., Matteucci F., Salucci P., Chiappini C., 2000, ApJ 539, 235
\bibitem[1996]{ryan96}Ryan S.G., Norris J.E., Beers T.C., 1996, ApJ 471, 254
\bibitem[1995]{ryde95}Ryder S. D., 1995, ApJ 444, 610
\bibitem[1997]{sack97}Sackett P., 1997, ApJ 483, 103
\bibitem[1955]{salp55}Salpeter E., 1955, ApJ 121, 161
\bibitem[1998]{saml98}Samland M., 1998, ApJ 496, 155
\bibitem[1997]{saml97}Samland M., Hensler G., Theis Ch., 1997, ApJ 476, 544
\bibitem[1986]{scal86}Scalo J.M., 1986, FCPhys 11, 1
\bibitem[1998]{scal98}Scalo J.M., 1998, in The Stellar Initial Mass Function, ed. G. Gilmore \& D.
Howell, ASP Conf. Ser., Vol. 142, 201
\bibitem[1992]{scha92}Schaller G., Schaerer D., Meynet G., Maeder A., 1992, A\&AS 96, 269
\bibitem[1959]{schm59}Schmidt M., 1959, ApJ 129, 243
\bibitem[1992]{shig92}Shigeyama T., Nomoto K., Yamahoka H., Thielemann F.-K., 1992, ApJ 386, L13
\bibitem[1988]{sned88}Sneden C., Crocker D.A., 1988, ApJ 335, 406
\bibitem[1991]{sned91}Sneden C., Gratton R.G., Crocker D.A., 1991, A\&A 246, 354
\bibitem[1994]{sned94}Sneden C., Preston G. W., McWilliam A., Searle L., 1994, ApJ 431, L27
\bibitem[1994]{stei94}Steinmetz M., Müller E., 1994, A\&A 281, L97
\bibitem[1999]{step99}Stephens A., 1999, AJ 117,1771
\bibitem[1975]{talb75}Talbot R.J., Arnett W.D., 1975, ApJ 197, 551
\bibitem[1994]{tamm94}Tamman G., Loefler W., Schroder A., 1994, ApJS 92, 487
\bibitem[1993]{thie93}Thielemann F.-K., Nomoto K., Hashimoto M., 1993, in Origin and Evolution of
the Elements, ed. N. Prantzos, E. Vangioni-Flam, \& M. Cassé (Cambridge: Cambridge
Univ. Press), 297
\bibitem[1996]{thie96}Thielemann F.-K., Nomoto K., Hashimoto M., 1996, ApJ 460, 408
\bibitem[1998]{thom98}Thomas D., Greggio L., Bender R., 1998, MNRAS 296, 119
\bibitem[1995]{timm95}Timmes F.X., Woosley S.E., Weaver T.A., 1995, ApJS 98, 617 (TWW1995)
\bibitem[1980]{tins80}Tinsley B.M., 1980, FCPh 5, 287
\bibitem[1986]{tomk86}Tomkin J., Sneden C., Lambert D.L., 1986, ApJ 302, 415
\bibitem[1988]{tosi88}Tosi M., 1988, A\&A 197, 47
\bibitem[1980]{twar80}Twarog B.A., 1980, ApJ 242, 242
\bibitem[1997]{vand97}van den Hoek L.B., Groenewegen M.A.T., 1997, A\&AS 123, 305
\bibitem[1999]{wakk99}Wakker B.P., Howk J.C., Savage B.D., van Woerden H., Tufte S.L., Schwarz U.J.,
Benjamin R., Reynolds R.J., Peletier R.F., Kalberla P.M.W., 1999, Nature 402,
388
\bibitem[1989]{whee89}Wheeler J.C., Sneden C., Truran J.W., 1989, ARA\&A 27, 289
\bibitem[1994]{woos94a}Woosley S.E., Eastman R.G., 1994, in Proc. Menorca Summer School on Supernovae,
ed. E. Bravo, R. Canal, J.M. Ibañez, J. Isern, 105
\bibitem[1994]{woos94b}Woosley S.E.,Weaver T.A., 1994, ApJ 423, 371
\bibitem[1995]{woos95}Woosley S.E., Weaver T.A., 1995, ApJS 101, 181 (WW1995)
\bibitem[1995]{wyse95}Wyse R.F.G, Gilmore G., 1995, AJ 110, 2771
\bibitem[1990]{zhao90}Zhao G., Magain P., 1990, A\&A 238, 242
\end{thebibliography}
\end{document}